\documentclass[a4paper,fleqn,usenatbib]{mnras}
\usepackage{newtxtext,newtxmath}

\usepackage[T1]{fontenc}
\usepackage{ae,aecompl}
\usepackage{graphicx}	
\usepackage{amsmath}	
\usepackage{amssymb}	
\usepackage{gensymb}
\usepackage{rotating} 
\usepackage{xcolor}
\title[KURVS: rotation curves \& dark matter at $z \sim 1.5 $]
{KURVS: The outer rotation curve shapes and dark matter fractions of $z \sim 1.5 $ star-forming galaxies.}

\author[A. Puglisi et al.]{Annagrazia Puglisi$^{1,2}$\thanks{E-mail: a.puglisi@soton.ac.uk (AP)},
Ugn\.{e} \,Dudzevi\v{c}i\={u}t\.{e}$^{3}$,
Mark Swinbank$^{1}$, 
Steven Gillman$^{4,5}$,
\newauthor
Alfred L. Tiley$^{1,6}$,
Richard G. Bower$^{1,7,8}$
Michele Cirasuolo$^{9}$,
Luca Cortese$^{6,10}$, 
\newauthor
Karl Glazebrook$^{11}$,
Chris Harrison$^{12}$,
Edo Ibar$^{13}$,
Juan Molina$^{14}$,
Danail Obreschkow$^{6,15}$,
\newauthor
Kyle A. Oman$^{7, 16}$,
Matthieu Schaller$^{17,18}$,
Francesco Shankar$^{2}$, 
and Ray M. Sharples$^{19}$
\\
\\
$^{1}$ Centre for Extragalactic Astronomy, Department of Physics, Durham University, South Road, Durham DH1 3LE, UK \\
$^{2}$ School of Physics and Astronomy, University of Southampton, Highfield SO17 1BJ, UK\\
$^{3}$ Max-Planck-Institut f\"ur Astronomie, K\"onigstuhl 17, D-69117, Heidelberg, Germany\\
$^{4}$ Cosmic Dawn Center (DAWN) \\
$^{5}$ DTU-Space, Technical University of Denmark, Elektrovej 327, DK-2800 Kgs. Lyngby, Denmark \\
$^{6}$ International Centre for Radio Astronomy Research, University of Western Australia, 35 Stirling Highway, Crawley, WA 6009, Australia\\
$^{7}$ Institute for Computational Cosmology, Durham University, South Road, Durham DH1 3LE, United Kingdom\\
$^{8}$ Institute for Data Science, Department of Physics, Durham University, South Road, Durham DH1 3LE, United Kingdom\\
$^{9}$ European Southern Observatory, Karl-Schwarzschild-Str 2, D-86748 Garching b. M\"unchen, Germany \\
$^{10}$ ARC Centre of Excellence for All Sky Astrophysics in 3 Dimensions (ASTRO 3D), Australia\\
$^{11}$ Centre for Astrophysics and Supercomputing, Swinburne University of Technology, PO Box 218, Hawthorn, VIC 3122, Australia\\
$^{12}$ School of Mathematics, Statistics and Physics, Newcastle University, Newcastle upon Tyne NE1 7RU, UK\\
$^{13}$ Instiuto de F\'isica y Astronom\'ia, Universidad de Valpara\'iso, Avda. Gran Breta\~na 1111, Valpara\'iso, Chile\\
$^{14}$ Department of Space, Earth and Environment, Chalmers University of Technology, Onsala Space Observatory, 439 92 Onsala, Sweden\\
$^{15}$ Australian Research Council Centre of Excellence for All-Sky Astrophysics, 44 Rosehill Street Redfern, NSW 2016, Australia\\
$^{16}$ Department of Physics, Durham University, South Road, Durham DH1 3LE, United Kingdom\\
$^{17}$ Lorentz Institute for Theoretical Physics, Leiden University, PO Box 9506, NL-2300 RA Leiden, The Netherlands\\
$^{18}$ Leiden Observatory, Leiden University, PO Box 9513, NL-2300 RA Leiden, The Netherlands\\
$^{19}$ Centre for Advanced Instrumentation, Department of Physics, Durham University, South Road, Durham DH1 3LE UK 
}
\date{Accepted XXX. Received YYY; in original form ZZZ}

\pubyear{2021}

\begin{document}
\label{firstpage}
\pagerange{\pageref{firstpage}--\pageref{lastpage}}
\maketitle

\begin{abstract}
We present first results from the KMOS Ultra-deep Rotation Velocity Survey (KURVS), aimed at studying the outer rotation curves shape and dark matter content of 22 star-forming galaxies at $z \sim 1.5$.
These galaxies represent `typical' star-forming discs at $z \sim 1.5$, being located within the star-forming main sequence and stellar mass-size relation with stellar masses $9.5 \leqslant$ log$(M_{\star}/\mathrm{M_{\odot}}) \leqslant 11.5$.
We use the spatially-resolved H$\alpha$ emission to extract individual rotation curves out to {4 times the effective radius, on average,} or $\sim 10-15$ kpc.
Most rotation curves are flat or rising between three and six disc scale radii. 
{Only three objects with dispersion-dominated dynamics ($v_{\rm rot}/\sigma_0 \sim 0.2$) have declining outer rotation curves at more than 5$\sigma$ significance.}
After accounting for seeing and pressure support, {the nine} rotation-dominated discs with $v_{\rm rot}/\sigma_0 \geqslant 1.5$ have average dark matter fractions of $50 \pm 20\%$ at the effective radius, similar to local discs.
Together with previous observations of star-forming galaxies at cosmic noon, our measurements suggest a trend of declining dark matter fraction with increasing stellar mass and stellar mass surface density at the effective radius.
Measurements of simulated EAGLE galaxies are in quantitative agreement with observations up to log$(M_{\star} R_{\rm eff}^{-2} /\mathrm{M_{\odot} kpc^{-2}}) \sim 9.2$, and over-predict the dark matter fraction of galaxies with higher mass surface densities by a factor of $\sim 3$.
We conclude that the dynamics of typical rotationally-supported discs at $z \sim 1.5$ is dominated by dark matter from effective radius scales, in broad agreement with cosmological models.
The tension with observations at high stellar mass surface density suggests that the prescriptions for baryonic processes occurring in the most massive galaxies (such as bulge growth and quenching) need to be reassessed. 
\end{abstract}


\section{Introduction}

Some of the first evidence for dark matter in galaxy evolution dates back to 1978, when observations demonstrated that stars in the outer regions of luminous spiral galaxies (at radii up to $\sim$50 kpc, far beyond the stellar disc) rotate faster than expected from the gravity due to the observed gas and stars. This implied that most of the mass in these outer regions must be invisible \citep{Bosma78, Rubin78}.
Since this first discovery, several other key measurements have converged to favour the existence of substantial amounts of dark matter within and outside of galaxies \citep[e.g.,][]{OstrikerPeebles73,Ostriker74,Frenk85}.
Cold dark matter (CDM) now provides the framework for structures formation and is essential for the formation and evolution of galaxies in the $\Lambda$CDM paradigm.
Most theoretical models in the last decades have thus included dark matter as a backbone for evolving galaxies in a cosmological context \citep[][see also \citealt{Vogelsberger20} for a recent review]{Vogelsberger14a, Schaye15, Dave16, Dave19, Dubois16, Dubois21, Springel18, Pillepich19}.
Cosmological simulations, built on $\Lambda$CDM theory, have had tremendous successes in recreating a Universe with many characteristics similar to those of our own \citep[][]{Genel14, Vogelsberger14b, Furlong17, Ferrero17}, and make specific predictions for the redshift evolution of baryonic and dark matter fractions within galaxies \citep[e.g.][]{Lovell18}.

Quantifying the mass and properties of baryons and dark matter haloes at early times is essential to better understand how galaxies assemble across cosmic history, and the main physical processes that shape their evolution.
Eventually, studying dark matter at $1 \lesssim z \lesssim 3 $, i..e cosmic noon, when galaxies are most efficiently forming stars and growing black holes at their centres \citep{MadauDickinson14}, can provide important constraints on the processes regulating its interaction with the baryonic matter.
Measuring baryonic and dark matter masses in the distant Universe is challenging as galaxies become gradually smaller, less luminous and more irregular at earlier cosmic times \citep[e.g. ][]{Conselice14}.
However, advances in integral-field spectroscopic techniques have allowed us to measure the kinematics of galaxies beyond $z \gtrsim 1$, thus opening up the field of observational studies of the total mass budget at early times.

{ Dynamical studies with integral-field unit (IFU) spectrographs and observations at moderate to high spatial resolution in the sub-millimetre} in the last decade have allowed us to characterise the kinematic properties of galaxies at cosmic noon and beyond \citep[e.g.,][]{Lelli18, Ubler18, Gillman19, Molina19, Sweet19, Kaasinen20, Fraternali21,HerreraCamus22}, also with the aid of gravitational lensing \citep[e.g., ][]{Rizzo20, Rizzo21}.
Kinematic surveys reveal that galaxies on the star-forming main sequence are predominantly rotating discs \citep{ForsterSchreiber09,ForsterSchreiber18,Wisnioski15, Wisnioski19, Stott16}.
These galaxies appear to be different than massive discs at $z \sim 0$, as turbulent motions (probed by the velocity dispersion in spatially-resolved spectra) provide a significant contribution to their dynamics \citep[e.g.][]{Weiner06a, Johnson18, Ubler19}.
The fraction of rotating discs increases significantly since $z \sim 3.5$ \citep{Turner17} and with increasing stellar mass at fixed redshift, suggesting that discs settle towards later times and that massive galaxies are more dynamically mature at a given cosmic epoch \citep[a phenomenon that has been referred to as ``kinematic downsizing'', e.g.][]{Kassin12, Simons17}.
To investigate the budget of visible and dark matter in disc galaxies and its evolution with redshift, dynamical surveys have exploited scaling relations such as the Tully-Fisher relation \citep{TullyFisher77, BellDeJong01}. 
This dynamical scaling relation compares the stellar mass \citep[or the total baryonic mass, if considering the Baryonic Tully-Fisher relation, e.g.][]{McGaugh00, Lelli16} to the rotation velocity, a tracer of the total dynamical mass, and hence provides information about the baryon and dark matter content of galaxies. 
A general consensus on the evolution of the Tully-Fisher relation as a function of redshift is still lacking, with some studies indicating mild or no evolution \citep{Kassin07, Miller12, Harrison17, Tiley19a, Gogate22} and others strong evolution \citep{Cresci09, Turner17b, Ubler17}.

Rotation curves describe the rotation velocity of the galaxy material as a function of galactocentric distance, and represent a more direct tool to probe the baryonic and dark matter content and radial distribution in and around galaxies. 
In the galaxy's central regions, stars and gas are expected to provide the dominant contribution to the rotation curve.
Far from the galactic centre, rotation curves are expected to be flat if dark matter becomes the dominant mass component of the galaxies' outer discs. 
Observations extending out to the edge of the stellar and gaseous disc ($\sim 10 - 15$ kpc) are thus required to constrain the dark matter halo mass.
IFU facilities at 8-meter class telescopes such as the Very Large Telescope have made possible to study rotation curves at high redshift. 
Recent studies using spatially-resolved observations of the H$\alpha$ emission line have suggested that some massive star-forming galaxies at $z \geqslant 1$ have declining rotation curves at large radii, potentially at odds with expectations from $\Lambda$CDM models \citep{Genzel17, Genzel20, Price21}.
It is still unclear how these results apply to the general star-forming population, and if dark matter fractions are a function of redshift, as also suggested by some Tully-Fisher evolution studies and statistical analyses of the total mass budget of galaxies \citep[e.g.][]{Wuyts16}, or other galaxy properties, such as the stellar mass.
For example, deep observations indicate that $z \sim 1$ star-forming galaxies at log$(M_{\star}/\mathrm{M_{\odot}}) \lesssim 10$, i.e. below the typical mass range probed by high-redshift surveys, show diverse shapes, ranging from rising to declining at large radii, and high dark matter fractions at scales of the stellar effective radius \citep{Bouche22}.
To understand if declining rotation curves and subsequently low dark matter fractions are the result of selection biases, and how rotation curve shapes and dark matter fractions are related to galaxy integrated properties and redshift, it is therefore essential to measure rotation curves in statistical samples covering a broad range in distant galaxy properties. 
However, dynamical surveys of $z \sim 1-3$ galaxies have typical integration times of 6-10 hrs \citep{FSW20} resulting in rotation curves of statistical samples that can be traced out to a few times the galaxy disc-scale radius, at best ($\sim R_{\rm 3D}$, equivalent to  $\sim$1.8 times the half-light radius or $\sim$7 kpc), and this is not sufficient for probing rotation curves at large radii where dark matter starts to provide dominant contribution to potential. 
Some studies rely on stacking to access statistical samples and trace average rotation curves out to $\sim$15 kpc \citep[e.g.][]{Lang17, Sharma21}, but it has been shown that stacking techniques are prone to systematics, and different assumptions can lead to differences in the implied dark matter fractions \citep{Tiley19}.
Therefore, deep observations of statistical samples of {\it individual} galaxies spanning a broad range of distant galaxies' properties are needed, probing the outer radii where dark matter is expected to provide the dominant contribution to the disc dynamics.

In this context, the KMOS Ultra-deep Rotational Velocity Survey (KURVS) is designed to push the limit of the K-band Multi Object Spectrograph (KMOS) to measure the shape of the rotation curves in a statistical sample of main-sequence galaxies at $z \sim 1.5$ on a galaxy-by-galaxy basis out to (and beyond) six times the disc scale radius ($\sim$10-16 kpc). 
This is achieved by exploiting the multiplex capabilities of KMOS \citep{Sharples13}, which allows us to target up to 24 galaxies simultaneously, to observe the spatially-resolved H$\alpha$ emission for up to $\sim 80- 100$ hours of integration per galaxy.
In this paper we describe the sample selection, observations and data reduction of KURVS. 
We then present measurements of the dynamical properties and outer rotation curve shapes of 22 KURVS galaxies in the Chandra Deep Field South (CDFS), each observed for $\approx$70 hours on source.
The spatially-resolved gas-phase metallicity properties of this sample have been presented in \citet{Gillman22}.
We use the \textsc{galpak$^{\rm 3D}$} parametric tool \citep{Bouche15} to correct the inner rotation curves for beam smearing, and measure dark matter fractions at the effective radius ($\sim$2-4 kpc). 
We then compare our measurements of the outer rotation curve shapes and dark matter content with typical discs in the local Universe, as well as with measurements of the dark matter fraction in other samples of star-forming galaxies at cosmic noon. 
To interpret our results in the $\Lambda$CDM context, we compare our results to model star-forming galaxies from the EAGLE cosmological simulation.

This paper is structured as follows. 
In Section \ref{Sect:KURVS} we describe the sample selection and properties of the sample in the context of the parent star-forming population at similar cosmic epochs. 
We further describe the observing strategy and data reduction process. 
In Section \ref{Sect:Analysis} we describe the derivation of resolved properties and kinematics of the subset of KURVS galaxies in CDFS, putting our measurements in the context of wider IFU surveys. 
In this section we also discuss measurements of the outer rotation curve shapes, the beam-smearing correction procedure, and measurements of the dark matter fraction at the effective radius. 
Additionally, we compare our measurements with previous results from the literature, as well as with simulations and local discs. 
Finally, we discuss our results in Section \ref{Sect:Discussion}.

Throughout this paper we assume a \citet{Chabrier} initial mass function and a standard $\Lambda$CDM cosmology ($H_{0} = 70$ km s$^{-1}$ Mpc$^{-1}$, $\Omega_{\rm m} = 0.3$, $\Omega_{\Lambda} = 0.7$).

\section{The KURVS survey}
\label{Sect:KURVS}

KURVS is an ESO Large Programme (ID 1102.B-0232) designed to study the spatially resolved dynamics of individual $z \sim 1.5$ galaxies at large radii with KMOS. 
In this section we provide an overview of the sample selection and its properties in the context of the parent star-forming population at similar redshift. 
We also describe the observations and key aspects of the data reduction for the subset of galaxies presented in this paper, corresponding to half of the sample. 
A similar observing strategy and data reduction will be applied to the full sample that will be presented in a forthcoming paper.

\subsection{Sample selection and galaxy integrated properties}
\label{Sect:Selection}

KURVS aims at obtaining ultra-deep spatially-resolved observations of the H$\alpha$+[NII]$_{6548, 6583}$ kinematics in star-forming galaxies at $z \sim 1.5$ by exploiting the multiplex capabilities of KMOS, which allows us to obtain integral field observations for up to 24 galaxies simultaneously.
Therefore, the target galaxies have been selected to be located within the $7''.2 \times 7''.2$ KMOS patrol field, in order to be observed within a single KMOS pointing. 
In addition, the targets have been selected to lie within the Cosmic Assembly Near-infrared Deep Extragalactic Legacy Survey \citep[CANDELS,][]{Grogin11} fields, to ensure access to high resolution multi-wavelength photometry from {\it HST}, as well as low-resolution multi-wavelength photometry from the ultra-violet to the near-infrared for the full sample. 
The final sample consists of 44 galaxies with existing H$\alpha$ detections split in two KMOS pointings, with 22/44 objects in the Chandra Deep Field South (CDFS) field and 22/44 objects in the Cosmological Evolution Survey (COSMOS) field. 
Out of these, 35 galaxies (20/22 in CDFS and 15/22 in COSMOS) have been selected from the KMOS Galaxy Evolution Survey \citep[KGES][]{Gillman20, Tiley21} among galaxies with pre-existing H$\alpha$ detections and spatially-resolved kinematics. 
The remaining two galaxies within CDFS and the seven COSMOS targets have been selected from the KMOS$^{\rm 3D}$ survey \citep{Wisnioski19} to fully populate the KMOS arms. One of the COSMOS sources is undetected due to an incorrect spectroscopic redshift, yielding a final sample of 43 galaxies.

KURVS galaxies have spectroscopic redshifts  $1.23 \leqslant z_{\rm spec} \leqslant 1. 71$, with a median redshift $z_{\rm spec} = 1.50$.
Integrated measurements of stellar masses, $M_{\star}$ and star formation rates, SFR, are fully described in \citet{Gillman20}, and we briefly report details of the adopted procedure below.
The same procedure has been applied to the galaxies not included in the original KGES sample. 
Stellar masses and star formation rates are measured by fitting the UV-to-near-IR spectral energy distribution (SED) from the CDFS and COSMOS photometric catalogues \citep{Guo13, Muzzin13} with \textsc{magphys} \citep{daCunha08, daCunha15}.
Galaxies in the KURVS sample span a stellar mass range log$(M_{\star}/ \mathrm{M_{\odot}}) = 9.2 - 11.5$, with a median log$(M_{\star}/ \mathrm{M_{\odot}}) = 10.2^{\rm + 0.5}_{\rm -0.4}$.
Star formation rates cover a range SFR$= 5 - 143 \ {\rm M_{\odot}} {\rm yr}^{-1}$, with a median SFR $= 22^{\rm + 19}_{\rm -9}  \ {\rm M_{\odot}}{\rm yr}^{-1}$. 
Here the uncertainties indicate the $16^{\rm th}$ to $84^{\rm th}$ percentile range of the distribution.
We obtain stellar continuum half-light radii, $R_{\rm eff}$, and axial ratios, $b/a$, from the \citet{vanderWel14} catalogue. 
These are based on {\it HST}-F125W imaging at $\lambda_{\rm obs} \sim 1.25 \ \mu$m if the redshift of the source is $z_{\rm spec} \leqslant 1.25$, and on F160W imaging at $\lambda_{\rm obs} \sim 1.6 \ \mu$m otherwise. The targets have half-light radii in the range $R_{\rm eff}$ = 1.1 - 14 kpc with a median $R_{\rm eff} = 3.7 ^{\rm + 1.9}_{\rm -1.2} $ kpc.
Inclinations for the stellar continuum are measured from the $b/a$ axial ratio, assuming a thick disc with intrinsic thickness $q_0 = 0.2$ \citep[see Eqn. 3 in ][and references therein]{Gillman20}. 
We measure inclinations for the star-forming disc on {\it HST}-F814W images sampling the rest-frame UV emission in our sources by adopting a consistent procedure to that applied on near-infrared images.
Measurements of these quantities for KURVS-CDFS galaxies discussed in this work are reported in Table \ref{tab1:KURVS_sample}.

To identify candidate active galactic nuclei (AGN) in the CDFS sample we use the same AGN identification scheme adopted for the full KGES sample, which is based on the integrated [NII]/H$\alpha$ ratio and emission-line width, as well as infrared colours and X-ray luminosities (see \citealt{Gillman22} for details on the procedure).
Of the 22 sources in the KURVS-CDFS sample, we identify two galaxies (KURVS-11 and KURVS-12) whose infrared colours indicate the presence of an AGN. KURVS-12 also has a X-ray counterpart and an intrinsic 0.5 -- 7.0 kev luminosity of $L_{\rm X-ray, int} = 5 \times 10^{42} \ \mathrm{erg \ s}^{-1}$ in the \citet{Luo17} catalogue, which  also supports the presence of an AGN. Three other galaxies in the sample have X-ray counterparts but luminosities that lie below the AGN threshold. 
These objects likely hosting an AGN do not show highly perturbed velocity or velocity dispersion maps, hence we do not exclude them from our analysis.

Owing to the adopted selection criteria mainly based on their projected sky distribution, KURVS galaxies are a representative subset of spectroscopically-selected $z \sim 1.5$ galaxies with log$(M_{\star}/M_{\odot}) \sim 10$. 
This is demonstrated in Figure \ref{fig:Selection, MS_MSize}, showing the distribution of KURVS with respect to KGES galaxies in the stellar mass versus star formation rate plane (left), and in the stellar mass versus half-light radius plane (right).
Figure \ref{fig:Selection, MS_MSize} also shows that the $\sim70 \mathrm{~per~cent}$ of KURVS galaxies are located within a factor of 4 of the main sequence at $z \sim 1.5$ \citep{Schreiber15} and within the scatter of the stellar mass-size relation for $z \sim 1.25 - 1.75$ discs \citep{vanderWel14}. 
Therefore, their star formation rate and structural properties suggest that KURVS galaxies are representative of $z \sim 1.5$ star-forming discs at log$(M_{\star}/M_{\odot}) \sim 10$ and log$({\rm sSFR}/\rm yr^{-1}) \sim -9$.

\begin{figure*}
\begin{center}
\centering
\includegraphics[scale=0.58]{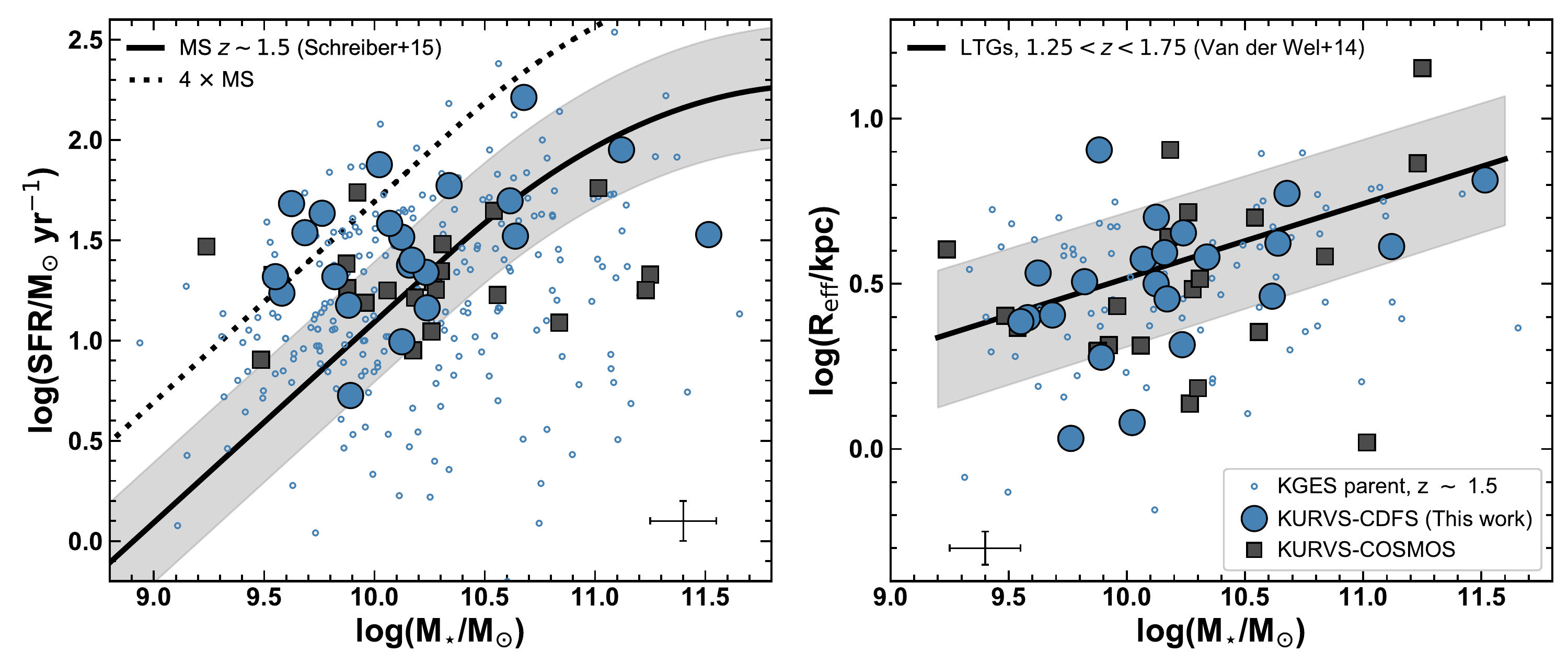}
\caption{ {\bf Left:} Star formation rate as a function of stellar mass for our sample. Here the star formation rate of each galaxy is normalised to the SFR of the main sequence at $z \sim 1.5$, to account for the redshift evolution of the main sequence normalisation \citep[e.g.,][]{Sargent12}.
The blue filled circles correspond to KURVS galaxies in the CDFS field discussed in this work. Grey filled squares indicate KURVS galaxies in COSMOS, which will be presented in a forthcoming paper. Blue hollow circles show the parent sample from the KGES survey \citep{Tiley21}. The median error bar for the KURVS and KGES points is shown in the bottom right.
The solid line and shaded area indicate the position of the main sequence and its $\pm 0.3$ dex scatter from \citet{Schreiber15} at $z = 1.5$, corresponding to the average redshift of the KURVS sample. The dashed line represents the factor of four threshold above which galaxies are usually classified as ``starbursts''. 
{\bf Right:} Stellar half-light radius as a function of stellar mass for the KURVS sample. The median error bar for KURVS and KGES galaxies is shown in the bottom left.
The black line and shaded area indicate the mass-size relation and its scatter at $1.25 < z < 1.75$ from \citet{vanderWel14}.
KURVS galaxies are located within the main-sequence and mass-size relations at the same redshift. 
Therefore, their star formation rate and structural properties suggest that KURVS galaxies represent typical star-forming systems at $z \sim 1.5$.}
\label{fig:Selection, MS_MSize}
\end{center}
\end{figure*}

\begin{table*}
\centering
\caption{Integrated properties of KURVS-CDFS galaxies.}
\label{tab1:KURVS_sample}
\begin{tabular}{cccccccccc} 
\hline
\hline
KURVS ID & CANDELS ID & RA & Dec. & $z_{\rm spec, H\alpha}$  &  log($M_{\star}$) & $SFR$ & R$_{\rm eff}$ & $i_{\rm \star}$ & $i_{\rm SFR}$ \\
 &  & hh:mm:ss & dd:mm:ss &  & log($\mathrm{M_{\odot}}$) & M$_{\odot}$/yr & kpc & deg & deg \\
(1) & (2) & (3) & (4) & (5) & (6) & (7) & (8) & (9) & (10) \\
\hline
1 & cdfs\_24904 & 03:32:17.35 & --27:53:52.83 & 1.359 & 9.76 & 36 & 1.1 $\pm$ 0.1 & 39 $\pm$ 5 & 44 $\pm$ 5 \\
2 & cdfs\_26404 & 03:32:15.00 & --27:53:02.37 & 1.360 & 9.88 & 12 & 8.0 $\pm$ 0.9 & 49 $\pm$ 6 & 48 $\pm$ 6 \\
3 & cdfs\_26954 & 03:32:22.10 & --27:52:44.95 & 1.541 & 10.64 & 34 & 4.2 $\pm$ 0.5 & 43 $\pm$ 5 & 42 $\pm$ 5 \\
4 & cdfs\_27318 & 03:32:20.18 & --27:52:38.34 & 1.552 & 10.34 & 62 & 3.8 $\pm$ 0.4 & 43 $\pm$ 5 & 53 $\pm$ 6 \\
5 & cdfs\_28138 & 03:32:13.78 & --27:52:02.73 & 1.518 & 10.12 & 33 & 3.2 $\pm$ 0.4 & 57 $\pm$ 7 & 53 $\pm$ 6 \\
6 & cdfs\_29207 & 03:32:14.05 & --27:51:24.40 & 1.221 & 10.61 & 35 & 2.9 $\pm$ 0.3 & 55 $\pm$ 6 & 63 $\pm$ 7 \\
7 & cdfs\_29589 & 03:32:11.23 & --27:51:07.10 & 1.518 & 10.24 & 14 & 4.5 $\pm$ 0.5 & 78 $\pm$ 9 & 78 $\pm$ 9 \\
8 & cdfs\_29831 & 03:32:06.83 & --27:50:55.37 & 1.540 & 9.58 & 18 & 2.5 $\pm$ 0.3 & 56 $\pm$ 6 & 68 $\pm$ 8 \\
9 & cdfs\_30267 & 03:32:21.52 & --27:50:40.53 & 1.540 & 10.12 & 10 & 5.0 $\pm$ 0.6 & 50 $\pm$ 6 & 48 $\pm$ 5 \\
10 & cdfs\_30450 & 03:32:29.92 & --27:50:31.91 & 1.390 & 9.89 & 4 & 1.9 $\pm$ 0.2 & 73 $\pm$ 8 & 72 $\pm$ 8 \\
11 & cdfs\_30557 & 03:32:34.03 & --27:50:28.82 & 1.384 & 10.68 & 142 & 5.9 $\pm$ 0.7 & 68 $\pm$ 8 & 73 $\pm$ 8 \\
12 & cdfs\_30561 & 03:32:31.54 & --27:50:28.68 & 1.613 & 11.52 & 38 & 6.5 $\pm$ 0.8 & 60 $\pm$ 7 & 56 $\pm$ 6 \\
13 & cdfs\_30732 & 03:32:12.50 & --27:50:20.59 & 1.334 & 9.62 & 39 & 3.4 $\pm$ 0.4 & 39 $\pm$ 5 & 42 $\pm$ 5 \\
14 & cdfs\_30865 & 03:32:37.37 & --27:50:13.62 & 1.389 & 9.68 & 30 & 2.5 $\pm$ 0.3 & 59 $\pm$ 7 & 50 $\pm$ 6 \\
15 & cdfs\_31127 & 03:32:16.94 & --27:50:04.06 & 1.613 & 10.07 & 43 & 3.8 $\pm$ 0.4 & 65 $\pm$ 7 & 38 $\pm$ 4 \\
16 & cdfs\_31671 & 03:32:37.10 & --27:49:40.94 & 1.569 & 10.16 & 25 & 3.9 $\pm$ 0.5 & 76 $\pm$ 9 & 73 $\pm$ 8 \\
17 & cdfs\_31721 & 03:32:11.12 & --27:49:38.41 & 1.353 & 9.55 & 17 & 2.4 $\pm$ 0.3 & 42 $\pm$ 5 & 42 $\pm$ 5 \\
18 & cdfs\_33014 & 03:32:32.64 & --27:48:48.72 & 1.341 & 9.82 & 17 & 3.2 $\pm$ 0.4 & 55 $\pm$ 6 & 51 $\pm$ 6 \\
19 & cdfs\_33246 & 03:32:19.79 & --27:48:39.12 & 1.357 & 10.23 & 18 & 2.1 $\pm$ 0.2 & 39 $\pm$ 5 & 42 $\pm$ 5 \\
20 & cdfs\_35213 & 03:32:25.03 & --27:47:18.18 & 1.356 & 10.02 & 63 & 1.2 $\pm$ 0.1 & 19 $\pm$ 2 & 27 $\pm$ 3 \\
21 & GS4\_10784 & 03:32:31.43 & --27:51:37.48 & 1.382 & 10.17 & 21 & 2.8 $\pm$ 0.3 & 54 $\pm$ 6 & 54 $\pm$ 6 \\
22 & GS4\_16960 & 03:32:37.74 & --27:50:00.39 & 1.618 & 11.12 & 101 & 4.1 $\pm$ 0.5 & 38 $\pm$ 4 & 38 $\pm$ 4 \\
\hline
\hline
\end{tabular}
\\[2mm] 
\begin{flushleft}
{\bf Note.} (1) Galaxy ID from KURVS; (2) galaxy CANDELS ID \citep{Grogin11}; 
(3) and (4) galaxy coordinates; 
(5) spectroscopic redshift from the H$\alpha$ emission line;
(6) stellar mass from \textsc{magphys}. The typical uncertainty associated with this quantity is $\pm$ 0.2 dex \citep{Mobasher15};
(7) star formation rate mass from \textsc{magphys} SED-fitting. The typical uncertainty associated with this quantity is $\pm$ 0.1 dex;
(8) effective radius for the stellar component from \citet{vanderWel14}; 
(9) inclination for the stellar disc from \citet{vanderWel14};
(10) inclination for the star-forming disc from {\it HST}-F814W images.
Quoted errors for the effective radius and inclination correspond to a typical 0.05 dex uncertainty following \cite{vanDerWel12}. 
\end{flushleft}
\end{table*}

\subsection{Observations and data reduction}

Observations for the KURVS-CDFS pointing were carried out between October 2018 and December 2019 with the $H$-band filter, to detect the H$\alpha$+[NII] emission lines at $z \sim 1.5$. 
For this subset of the sample, 22 out of the 24 KMOS arms have been allocated to target galaxies, whilst one arm has been allocated to a star, in order to monitor the point spread function (PSF) of the observations, and to allow accurate centering of the individual frames during the data reduction process (see below).
Observations were carried out with an ABAABA observing scheme, where ``A" represents the science exposure and ``B" the sky frame, respectively. Each individual exposure lasted 600 s. A total of 418 frames have been considered to reconstruct the final science cubes, while 73 frames have been discarded due to poor seeing conditions to avoid degrading the spatial resolution of the science cubes. These ``poor seeing'' frames have been derived by inspection of the PSF of the standard star observed simultaneously with the science targets.
The total exposure time on-source for KURVS-CDFS observations is 69.7 hours, with an average seeing full width half maximum (FWHM) of 0.$''$57.

Data-cubes for each individual KMOS frame have been reconstructed using the ESO Recipe Execution Tool \citep[\textsc{esorex}, ][]{ESOrex} pipeline, which performs standard dark, flat and wavelength calibrations, and produces a 0.2$''$ spaxel data-cube for each frame. 
After the cube reconstruction, the sky has been subtracted from each frame. 
Firstly, \textsc{esorex} has been used to perform a simple A-B subtraction. 
Subsequently, the Zurich Atmospheric Purge tool \citep[\textsc{zap}, ][]{Soto16} has been applied to each sky-subtracted cube to remove residual contamination from the sky.
The frames have been flux-calibrated by using corresponding observations of the standard star taken with the target galaxies observations. 
Finally, to produce the stacked science cubes the calibrated frames have been centered using the position of the standard star. 

\begin{figure*}
\begin{center}
\centering
\includegraphics[scale=0.3]{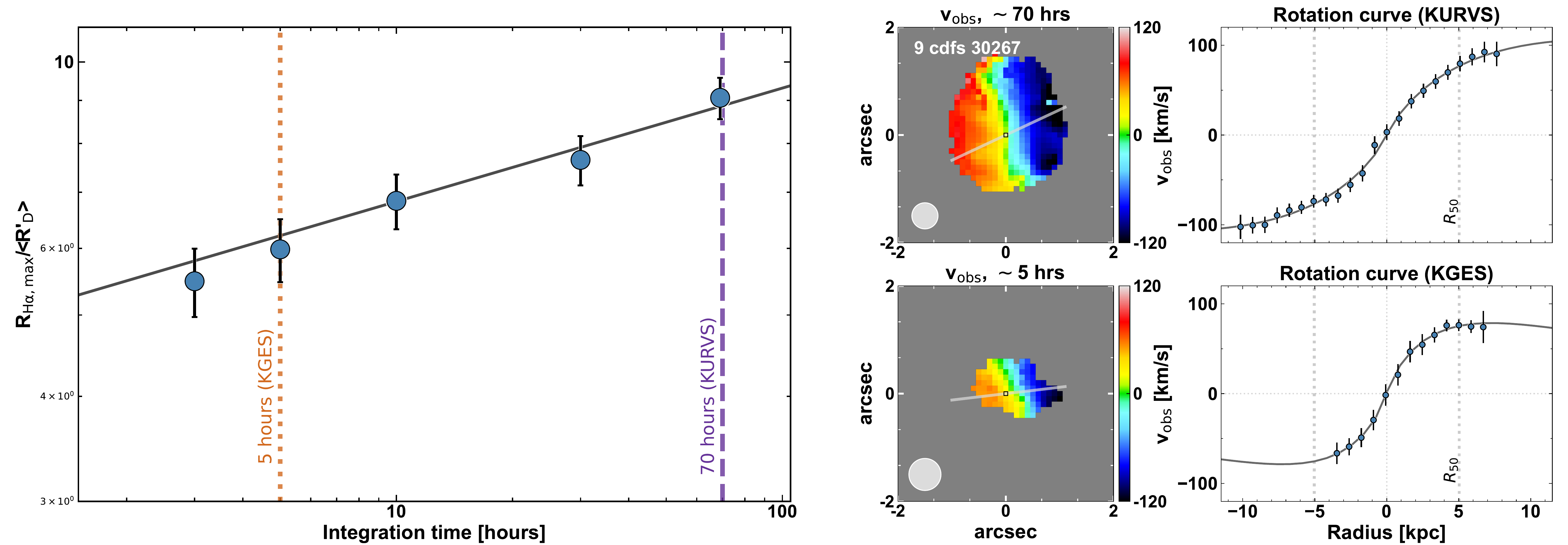}
\caption{ {\bf Left:} Average radial extent of the H$\alpha$ emission in KURVS galaxies as a function of integration time. 
Filled circles indicate the median extent of the H$\alpha$ emission, derived as the square root of the area in which the H$\alpha$ emission is detected with S/N$_{\rm H\alpha} \geqslant 5$. The errorbars highlight the 1$\sigma$ scatter of the distribution. 
Coloured vertical lines highlight the typical integration time of the parent sample from the KGES survey (dark orange) and KURVS (dark purple).
For 70 hours of on-source integration, corresponding to the integration time in the KURVS survey, we robustly sample galaxy rotation curves out to $\sim R_{\rm 6D}$ for the average system.
{\bf Right:} Example of the velocity field and rotation curve from KURVS (top row) and KGES (bottom row) observations for the same galaxy. 
These are measured from KMOS data-cubes  with a total on-source exposure time of $\sim$70 and $\sim$5 hours, respectively, as described in Section \ref{Sect:Analysis}.
The deep observations from KURVS allow us to measure high-quality velocity maps and rotation curves extending further out into the galaxy disc, as well as to derive the dynamical center with higher accuracy than pre-existent, shallower observations from KGES.}
\label{fig:Integration_OB_KURVS_vs_KGES}
\end{center}
\end{figure*}

The long integration of KURVS observations is critical to probe the outer regions of galactic discs at high-redshift.
This is demonstrated in Figure \ref{fig:Integration_OB_KURVS_vs_KGES}, showing the average radial extent of the H$\alpha$ emission as a function of integration time, and a comparison between the velocity field and rotation curves obtained from KURVS and KGES observations, the latter having a total on-source exposure time of $\sim 5$ hours, on average \citep{Tiley21}.
KGES observations allow us to recover the inner velocity field and velocity gradient of $z \sim 1.5$ galaxies, as exemplified by the right panel of Figure \ref{fig:Integration_OB_KURVS_vs_KGES}.
KGES observations slightly overestimate the circular velocity at the effective radius of KURVS-CDFS galaxies by 0.2 dex, on average, and show no significant systematics for measurements of the velocity dispersion when this is measured consistently as the median of the full velocity dispersion profile in both data-sets.
At the same time, kinematic parameters measured from KGES observations are associated with larger error bars, and overall KURVS and KGES measurements are consistent within the 1$\sigma$ uncertainties.
However, shallow observations from KGES require extrapolation of the rotation curve at large radii, and this results in significant uncertainties in measurements of the outer disc kinematics.
In fact, our previous analysis of the large-scale dynamics of $z \sim 1.5$ star-forming galaxies was based on stacking which allowed us to improve significantly the outer rotation curve signal-to-noise ratio \citep[modulo systematics associated with the normalisation of individual rotation curves, ][]{Tiley19}.

\section{Analysis}
\label{Sect:Analysis}

\subsection{Derivation of the emission line maps}

The systemic redshift of each source has been measured from one-dimensional spectra obtained by collapsing the cube in a pseudo-circular aperture with a $1.''2$ diameter, to maximise the signal-to-noise of the H$\alpha$ emission, S/N$_{\rm H\alpha}$. 
The H$\alpha$+[NII]$_{6548, 6583}$ complex and underlying continuum emission in these spectra has been modelled using a $\chi^2$ minimization procedure that uses three Gaussian profiles and a constant function. 
To account for increased noise near the sky emission lines, a weighting scheme corresponding to 1/$\mathrm{sky}^2$ has been applied, where `sky' is the sky spectrum from the KMOS data reduction pipeline. 
The three emission lines have a common width and a fixed relative position. 
Furthermore, the [NII]$_{6583}/$[NII]$_{6548}$ ratio is fixed to the theoretical value of 2.96.
The width and position of the emission line complex, as well as the normalisation of the continuum, are free parameters of the fit. 
Systemic redshifts measured on KURVS spectra are in excellent agreement with pre-existent measurements from KGES (within 1 per cent).

Each spaxel was re-sampled from the native $0.''2$ spaxel$^{-1}$ resolution to a $0.''1$ spaxel$^{-1}$ scale, conserving the flux in each slice during the process.
An adaptive binning procedure has been applied to the cubes, similarly to that applied in the analysis of data in the KROSS and KGES surveys \citep{Stott16, Tiley19, Tiley21}. 
For each spaxel, the flux is averaged in an increasing number of spaxels until S/N$_{\rm H\alpha} \geq 5$. 
Initially each spaxel is collapsed in a $\pm 1$ spaxel radius (resulting in a $3 \times 3$ spaxel). If the S/N$_{\rm H\alpha}$ of the resulting spaxel is below the S/N$_{\rm H\alpha} = 5$ threshold, the collapsing radius is iteratively increased by one pixel up to a maximal bin radius of 4 (corresponding to a  $9 \times 9$ {spaxel}) until S/N$_{\rm H\alpha} \geq 5$.
If the S/N$_{\rm H\alpha}$ is still below the threshold, the H$\alpha$ emission at that spaxel is considered undetected and the corresponding spaxel is masked in the final data cube.

\begin{figure*}
\begin{center}
\centering
\includegraphics[scale=0.34]{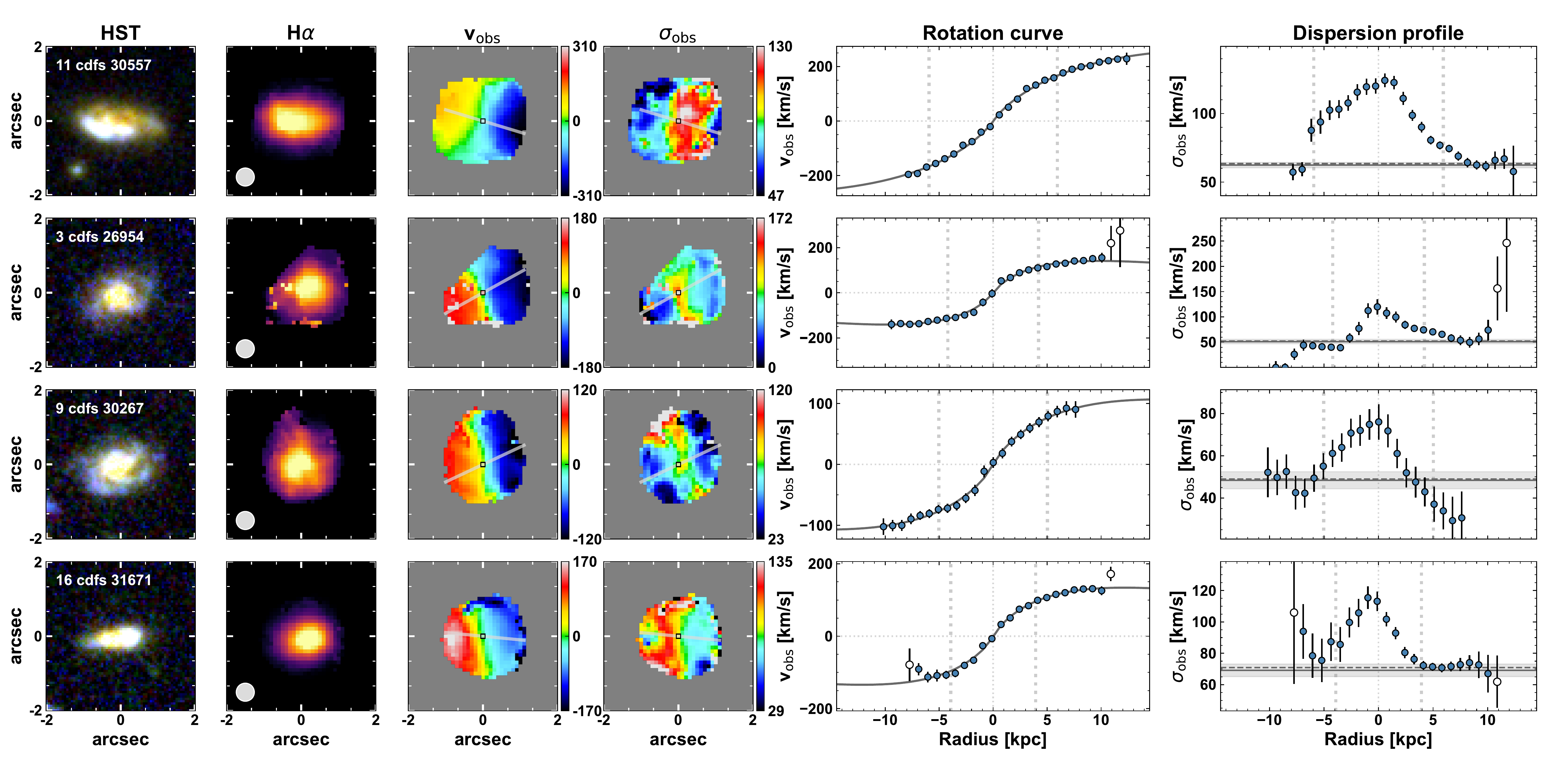}
\caption{The KURVS data set. From left to right, the panels show the {\it HST} $B_{435}- z_{850}-H_{160}$ composite image, the H$\alpha$ flux map, the line-of-sight velocity ($v_{\rm obs}$) and velocity dispersion ($\sigma_{\rm obs}$) maps, and the observed H$\alpha$ rotation curve and velocity dispersion profile. 
These are extracted from the KURVS velocity and velocity dispersion maps along the kinematic major axis, indicated with a grey solid line in each map. 
{Rotation curves are plotted such that the velocity gradient is always positive.
Note that the colour scale in the velocity and velocity dispersion maps might differ from the range of velocities of the one-dimensional profiles since it is optimised to the range of velocities and velocity dispersions measured in two-dimensional data.}
The size of the PSF of KURVS observations is indicated by a grey filled circle in the bottom left of each H$\alpha$ flux map. 
The white filled circles in the rotation curves and dispersion profiles indicate pixels that have been clipped due to either a large contamination from sky lines or broad components (see text for details).
{Here the thick dotted vertical lines mark the half-light radius measured from {\it HST} observations ($R_{\rm 50}$) for reference.}
The black solid curve in each rotation curve plot shows the best-fit exponential disc model to the data. 
For the dispersion profiles, the dotted grey and solid grey horizontal lines represent the observed and beam smearing corrected measure of the intrinsic dispersion, $\sigma_{0}$. The grey shaded area highlights the 1$\sigma$ error of this quantity. 
The depth of KURVS observations allows us to probe the rotation curves of $z \sim 1.5$ star-forming galaxies out to $\gtrsim
$ 10 kpc, well beyond the stellar effective radius, hence providing important observational constraints on the amount of baryonic and dark matter in their outer discs.}
\label{fig:kurvs}
\end{center}
\end{figure*}

The H$\alpha$+[NII]$_{6548, 6583}$ spatially-resolved emission is modelled by applying the same procedure used to fit the one-dimensional spectra to each spaxel in the binned data cube. 
To minimise contamination from any residual sky lines, the width of the emission lines is imposed to be wider than that of the sky lines. This is particularly relevant in regions with low S/N$_{\rm H\alpha}$ such as the outer regions of the galaxy disc. 
Maps for the H$\alpha$ flux, [NII]$_{6583}$ flux, observed line-of-sight velocity and observed velocity dispersion are constructed from the best-fit spatially-resolved model to the observed emission.
For this, the integral below the H$\alpha$ component, the integral of the [NII]$_{6583}$ emission line, the position of the H$\alpha$ emission line centroid relative to the systemic emission line centroid from the integrated spectrum, and the width of the emission lines are considered at each spaxel respectively.
Finally, the velocity dispersion is corrected for the instrumental broadening measured from the width of the sky emission lines. Uncertainties associated with each map are the 1$\sigma$ errors obtained from the $\chi^2$ minimisation routine. 
At this stage, additional shifts have been applied to recentre the spatially-resolved maps to account for the different morphology of the continuum and H$\alpha$ emission due to the presence of star-forming clumps and/or differential dust attenuation effects. 
The additional shifts are computed by fitting a two-dimensional Gaussian profile to the H$\alpha$ flux maps. 
The average shift along the {\it x}-axis is $x_{\rm shift} \sim -0.''03$, while along the {\it y}-axis is $y_{\rm shift} \sim -0.''04$. 
Figure~\ref{fig:kurvs} shows an example of the H$\alpha$ flux map, the observed line-of-sight velocity and observed velocity dispersion maps for a subset of the KURVS-CDFS galaxies. The full KURVS-CDFS sample is shown in Appendix \ref{A:KURVS data}.

\subsection{Kinematic position angles, rotation curves and velocity dispersion profiles}
\label{Sect:RCs_SCs_extraction}

To extract rotation curves and velocity dispersion profiles and hence characterise the dynamic properties of the sample, we must identify the kinematic major axis of a galaxy corresponding to the axis that maximises the velocity gradient in the velocity map.
We measure the position angle of the kinematic major axis by rotating the observed velocity map in one degree increments. 
At each step we measure the median velocity in a pseudo-slit of 5 pixels (i.e. 0.$''$5) radius along the centre of the map, derived as discussed in the previous section.
To minimise the impact of noise on measurements of the position angle, we smooth the velocity gradient as a function of angle curve.
We then define the position angle of the kinematic major axis as the average between the angle that maximises the velocity gradient along the pseudo-slit and the angle that minimises it plus 90\textdegree. 
{We visually inspect the data to confirm that the average between the two independent measurements of the position angle corresponds to the direction of the maximum velocity gradient in the velocity maps. The measurements agree within $\pm 5$\textdegree \ with few notable exceptions corresponding to sources in which the velocity gradient is not well defined as a result of low inclination and/or a perturbed velocity field. This suggests that our measurements of the position angle are robust against deviations from circular motions such as inflows or outflows, which are likely to occur along the minor axis. }

We extract the rotation curve along the kinematic major axis as the median velocity at each (radial) pixel within a pseudo-slit of 3 pixels (i.e. 0.$''$3) radius, roughly corresponding to the 1$\sigma$ width of the seeing in KURVS observations ($\sim 0.''25$).
To measure errors on the extracted rotation curve, we vary the observed velocity at each pixel within the pseudo-slit within the 1$\sigma$ errors 1000 times and we take the standard deviation of the simulated velocity values at each pixel.
We apply a similar procedure to the velocity dispersion maps to extract the velocity dispersion profile along the kinematic major axis. 
The last two panels of Figure \ref{fig:kurvs} show an example of the extracted rotation curves and velocity dispersion profiles for four of the KURVS-CDFS galaxies.
The full sample is shown in Appendix~\ref{A:KURVS data}.

\subsection{Kinematic parameters}
\label{Sec:kin_pars}

\subsubsection{Rotational velocities}

In order to quantify the amount of ordered motions in each galaxy, we measure the rotation velocity at different radii.
We measure observed rotational velocities in our galaxies from best-fit models to the rotation curves derived in the previous section. 
This is to minimise the impact of noise and to extrapolate the measurements where the data do not extend far out enough in the galaxy disc.
To parametrise each one-dimensional rotation curve we fit it with an exponential disc model \citep{Freeman70} of the form:
\begin{equation}
(v(r) - v_{\rm off})^2 = \frac{(r-r_{\rm off})^2\pi G \mu_{0}}{h}(I_0K_0 - I_1K_1),
\end{equation}
where $\mu_0$ and $h$ are, respectively, the peak mass surface density and the disc scale radius, and $I_{\rm n}$, $K_{\rm n}$ are the Bessel functions evaluated at 0.5$r/h$.  We also allow for a systematic radial and velocity offsets, $r_{\rm off}$ and $v_{\rm off}$, respectively. 
While a razor thin disc model may not be a good approximation for high-redshift star-forming galaxies, which mostly present thick and turbulent discs \citep[][and references therein]{FSW20}, the \cite{Freeman70} model provides a good fit to the observed rotation curves, and it is used here only to interpolate the observational data-points and for extrapolating the rotation curve at large radii. 

We measure the observed rotation velocity at different radii from the best-fit, centered exponential disc model. 
To account for the impact of the seeing, we measure `KMOS radii' $R'$ by convolving the radius measured on {\it HST} observations with the 1$\sigma$-width of the seeing in KURVS observations, following \citet{Tiley19}.
To measure 1$\sigma$ uncertainties on the observed rotation velocity, we resample each data-point in the observed rotation curve and the relevant radii 1000 times within the 1$\sigma$ errors, we fit each of these `mock' rotation curves and we measure `mock' observed rotational velocities. Finally, we consider as 1$\sigma$ error the standard deviation of these `mock' rotation velocity values.
We note that measurements of the rotation velocity are extrapolated from the best-fitting model when the observed data do not extend far out into the disc. 
For most sources this corresponds to a small extrapolation, and a comparison with the position-velocity diagrams show that the best-fitting model agree very well with the observed kinematics up to very large radii (see Figure \ref{fig:kurvs_pv} in Appendix \ref{A:KURVS data}).
After visually inspecting the rotation curves and position-velocity diagrams, we clip a small number of non-physical pixels in six objects, where these are either affected by residual contamination from sky lines or emission from broad lines. 

To obtain intrinsic measurements of rotational velocities and intrinsic velocity dispersion in the ``inner disc'', i.e. within 3.4 times the disc scale radius $R'_{\rm 3.4D}$ or 2 times the effective radius, we apply a beam-smearing correction by using the prescriptions provided by \citet{Johnson18}. 
Since the impact of beam smearing decreases as a function of galacto-centric radius \citep[see e.g. figure~2 in][]{Johnson18}, we assume that the beam smearing correction is negligible in the outer disc, and we do not apply beam smearing corrections for kinematic parameters computed beyond $R'_{\rm 3.4D}$.
We further correct the rotation velocity for inclination by considering the inclination measured on {\it HST}-F814W images sampling the rest-frame UV emission of our sources, and hence the star formation rate similarly to the H$\alpha$ emission (see Section \ref{Sect:Selection} and Table \ref{tab1:KURVS_sample}).

\subsubsection{Ionised gas velocity dispersions}
\label{Subsec:sigma_0}

To classify the kinematics of galaxies we also need to measure the velocity dispersion, which allows us to quantify the disordered motions in the disc.
We use the velocity dispersion profile to measure the observed velocity dispersion, $\sigma_{\rm 0, obs}$. 
We measure this quantity as the weighted mean of at least 3 pixels with S/N$\geqslant 3$ beyond either $R'_{\rm 3.4D}$ or 2.2 times the disc scale radius, $R'_{\rm 2.2D}$. 
This is a compromise between considering the outermost pixels to minimise the impact of beam smearing, and averaging out a sufficient number of pixels to minimise the impact of noise.
{For two galaxies (KURVS-2 and KURVS-12) the measurements are not sufficiently extended and/or less than 3 pixels are available with sufficient S/N beyond $R'_{\rm 2.2D}$. In these cases we compute $\sigma_{\rm 0, obs}$ as the weighted mean of the full velocity dispersion profile.}
We finally compute the intrinsic velocity dispersion $\sigma_{\rm 0}$ by using the beam smearing corrections {reported in Table B1 of \citet{Johnson18}, which are based on the radial extent at which the velocity dispersion is measured and the velocity gradient of the galaxy}.

{We tested the impact of the radial extent at which the velocity dispersion is measured by comparing measurements of the velocity dispersion in the outer disc to those derived from the full velocity dispersion profile in galaxies with velocity dispersion profiles extending beyond $R'_{\rm 2.2D}$. 
The two measurements of velocity dispersion are in good agreement when beam smearing corrections are applied, and we measure an average ratio of $1.13 \pm 0.26$. This suggests that no significant systematics are introduced when the velocity dispersion is measured from the full velocity dispersion profile and beam smearing corrections are properly accounted for.}

\subsection{Rotational support}
\label{Subsec:Rot_support}

The ratio between the rotation velocity $v_{\rm rot}$ and the velocity dispersion $\sigma_{0}$ quantifies the balance between circular and turbulent motions in a galaxy, hence provides an indication of its disc properties.
In the left panel of Figure \ref{fig:Dynamics, vsigma_sigmaz} we plot $v_{\rm rot}/\sigma_{0}$ as a function of the stellar mass for KURVS-CDFS galaxies. 
Here we measure the rotation velocity as the inclination-corrected velocity at 6 times the disc scale radius, $R'_{\rm 6D}$, on the rotation curve, and the intrinsic velocity dispersion $\sigma_{0}$ as described in Section \ref{Subsec:sigma_0}. 
As a reference, we also add measurements for the KGES parent sample \citep{Tiley21}.
We find $v_{\rm rot}/\sigma_{0} = 1.6 \pm 0.2 $ with a 16$^{\rm th}$ - 84$^{\rm th}$ percentile range of 1.0 -- 2.0.
Figure \ref{fig:Dynamics, vsigma_sigmaz} shows that KURVS-CDFS galaxies have $v_{\rm rot}/\sigma_{0}$ properties representative of the KGES parent sample. 
All but three KURVS-CDFS galaxies have $v_{\rm rot}/\sigma_{0} \geqslant 1$, suggesting that the majority of galaxies in our sample are rotationally supported. 
{We use a threshold of $v_{\rm rot}/\sigma_{0} \geqslant 1.5$ to identify rotationally-supported galaxies in our sample.
This is intermediate between the $v_{\rm rot}/\sigma_{0} \geqslant 1$ cut typically applied in large dynamical surveys at high redshift \citep[e.g.][]{Genzel06, Wisnioski15, Wisnioski19, Johnson18} and a more conservative cut of $v_{\rm rot}/\sigma_{0} \geqslant 3$ which selects a strictly rotation-dominated, ``discy’’ sub-sample \citep{Tiley19a, Tiley21}.}
Using this value, we find a disc fraction $f_{\rm disc} = 50 \pm 20 \%$ \footnote{{We find $f_{\rm disc} = 86 \pm 30 \%$ and $f_{\rm disc} = 10 \pm 7 \%$ when using a $v_{\rm rot}/\sigma_{0}$ cut of 1 and 3 respectively}}, which is consistent with the fraction of discs measured in large integral field surveys in a similar stellar mass and redshift range \citep{Kassin12, Simons16, Simons17, Johnson18, Wisnioski19, FSW20, Tiley21}.
{A visual inspection of the velocity and velocity dispersion maps, and one-dimensional velocity and velocity dispersion profiles confirms that the cut of $v_{\rm rot}/\sigma_{0} \geqslant 1.5$ allows us to select galaxies that have regular and rotationally-supported dynamics, line-of-sight velocity dispersion that peaks in the central regions and disc-like morphologies, hence representative of star-forming discs at $z \sim 1.5$.}

\begin{figure*}
\begin{center}
\centering
\includegraphics[scale=0.55]{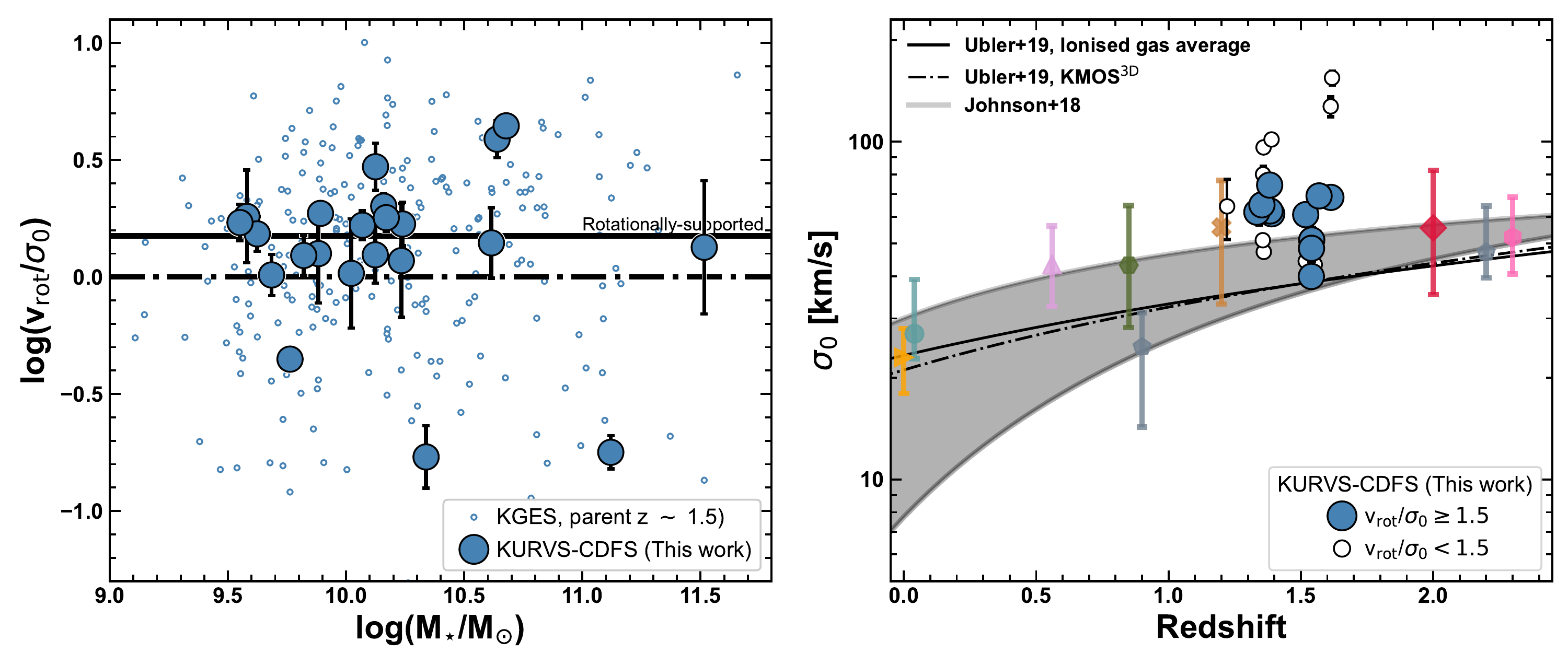}
\caption{{\bf Left:} Ratio between inclination corrected rotation velocity ($v_{\rm rot}$) and intrinsic velocity dispersion ($\sigma_{\rm 0}$) as a function of the stellar mass. 
The horizontal dash-dotted line marks the ratio $v_{\rm rot}$/$\sigma_{\rm 0} = 1$, which has been used to distinguish ``dispersion dominated'' ($v_{\rm rot}$/$\sigma_{\rm 0} < 1$) and ``rotation dominated'' ($v_{\rm rot}$/$\sigma_{\rm 0} > 1$) galaxies in the literature \citep[e.g.,][]{Wisnioski15, Johnson18}.
The horizontal solid line shows our conservative threshold $v_{\rm rot}$/$\sigma_{\rm 0} = 1.5$ to identify rotationally-supported galaxies in KURVS. 
The fraction of rotationally-supported KURVS-CDFS galaxies is consistent with measurements of the discs fraction from KGES in a similar redshift and stellar mass range \citep{Tiley21}. This is also consistent with KMOS$^{\rm 3D}$ measurements when applying the same $v_{\rm rot}$/$\sigma_{\rm 0} > 1$ cut \citep{Wisnioski19}.
{\bf Right:} Intrinsic velocity dispersion ($\sigma_{\rm 0}$) as a function of redshift.
Blue filled large circles and white small circles represent rotationally-supported and dispersion-dominated KURVS-CDFS galaxies, respectively. 
The coloured symbols are literature measurements for main-sequence star-forming galaxies at different redshifts
{(cyan circle: SAMI, \citealt{Bryant15};
orange triangle: GHASP, \citealt{Epinat10}; lilac triangle: MUSE, \citealt{Swinbank17}; dark grey pentagons: KMOS$^{\rm 3D}$, \citealt{Wisnioski15, Wisnioski19}; dark green hexagon: KROSS, \citealt{Johnson18}; dark yellow cross: MASSIV, \citealt{Epinat12}; red diamond: SIGMA, \citealt{Simons16}; pink hexagon: SINS, \citealt{Cresci09})}.
The solid line corresponds to the fit to literature measurements of the ionised gas velocity dispersion, and the dot-dashed line is the best-fit trend from KMOS$^{\rm 3D}$ \citep{Ubler19}. 
{The grey shaded area represent the redshift evolution of $\sigma_{0}$ for a Toomre disc instability toy model with log$(M_{\star}/[{\rm M_{\odot}}]) = 9.8 - 10.8$ \citep{Johnson18}, which corresponds to the $16^{\rm th} - 84^{\rm th}$ percentile range of the stellar mass distribution of the KURVS-CDFS sample (see Sect. \ref{Sect:Selection}).}
Rotationally-supported KURVS-CDFS galaxies follow the $\sigma_{0}-z$ trend highlighted by previous studies.
Overall, the integrated dynamical properties of KURVS-CDFS galaxies suggest that these are typical $z \sim 1.5$ star-forming discs at log$(M_{\star}/[{\rm M_{\odot}}]) \sim 10.2$.}
\label{fig:Dynamics, vsigma_sigmaz}
\end{center}
\end{figure*}

\subsection{Redshift evolution of the velocity dispersion}
\label{Subsec:sigma}

The right panel of Figure \ref{fig:Dynamics, vsigma_sigmaz} shows the intrinsic velocity dispersion as a function of redshift for KURVS-CDFS galaxies and a compilation of measurements from the literature. 
This figure shows that our sample displays velocity dispersions that are overall consistent with the trend expected from the redshift evolution of this quantity.
At the same time, KURVS-CDFS galaxies have higher $\sigma_0$ than the KMOS$^{\rm 3D}$ and KROSS trends \citep{Johnson18, Ubler19}.
However, we note that these trends are derived for ``kinematic samples'' including only rotation-dominated discs, as opposed to our observations which do not apply any a-priori cut on the rotational support properties of galaxies.

We measure $\sigma_{\rm 0} = 69 \pm 6$ km s$^{-1}$ with a $48 - 90$ km s$^{-1}$ 16$^{\rm th}$ - 84$^{\rm th}$ interquartile range. 
This is $\sim$1.5$\times$ higher than the velocity dispersion measured in galaxies from the KGES survey at similar redshifts \citep{Tiley21}. 
However, this difference can again be explained considering that \citet{Tiley21} measurements are restricted to a ``kinematic sample'' with main-sequence like SFR, spatially resolved H$\alpha$ emission up to $R'_{\rm 2.2D}$ and no signs of AGN emission, which is biased against sources with high velocity dispersion.
Indeed, the full KGES sample has an average velocity dispersion of $75 \pm 4$ km s$^{-1}$ with a $38 - 101$ km s$^{-1}$ 16$^{\rm th}$ -- 84$^{\rm th}$ interquartile range, and this is fully consistent with measurements of KURVS-CDFS galaxies.
We therefore conclude that KURVS-CDFS galaxies have velocity dispersions overall consistent with the star-forming population at $z \sim 1.5$. 
The higher values than the average can be explained by a combination of the sample selection, which does not include any a-priori cut on the dynamical properties of the galaxies, as well as by intrinsic variations within the galaxy population, showing substantial scatter at a given redshift \citep[e.g.,][]{Johnson18, Ubler19}.

\begin{table}
\centering
\caption{Kinematic properties of KURVS-CDFS galaxies.}
\label{tab2:kinematics}
\begin{tabular}{cccccc} 
\hline
\hline
KURVS ID & R$_{\rm H\alpha, max}$  & $\sigma_0$ & $v_{\rm rot}/\sigma_0$ & $t$  \\
 & kpc & km s$^{-1}$  & km s$^{-1}$ &  &  \\
  (1) & (2) & (3) & (4)  & (5) \\
\hline
1 & 10.7 & 96 $\pm$ 2 & 0.4 $\pm$ 0.1 & 1.05 $\pm$ 0.03 & \\
2 & 12.1 & 47 $\pm$ 1 & 1.3 $\pm$ 0.3 & 1.13 $\pm$ 0.12 & \\
3 & 11.7 & 51 $\pm$ 4 & 3.9 $\pm$ 0.1 & 0.92 $\pm$ 0.03 & \\
4 & 12.0 & 43 $\pm$ 2 & 0.2 $\pm$ 0.1 & 0.71 $\pm$ 0.05 & \\
5 & 12.1 & 44 $\pm$ 3 & 1.2 $\pm$ 0.2 & 1.19 $\pm$ 0.07 & \\
6 & 8.1 & 64 $\pm$ 13 & 1.4 $\pm$ 0.2 & 1.13 $\pm$ 0.10 & \\
7 & 11.4 & 61 $\pm$ 5 & 1.7 $\pm$ 0.2 & 0.98 $\pm$ 0.04 & \\
8 & 11.0 & 40 $\pm$ 3 & 1.8 $\pm$ 0.4 & 0.91 $\pm$ 0.03 & \\
9 & 10.2 & 48 $\pm$ 4 & 3.0 $\pm$ 0.3 & 1.04 $\pm$ 0.06 & \\
10$^{\star}$ & 15.2 & 61 $\pm$ 2 & 1.9 $\pm$ 0.1 & 0.99 $\pm$ 0.02 & \\
11 & 12.4 & 62 $\pm$ 2 & 4.4 $\pm$ 0.2 & 1.15 $\pm$ 0.04 & \\
12 & 11.2 & 127 $\pm$ 9 & 1.3 $\pm$ 0.3 & 0.67 $\pm$ 0.21 & \\
13 & 12.1 & 62 $\pm$ 4 & 1.5 $\pm$ 0.1 & 1.11 $\pm$ 0.05 & \\
14 & 9.8 & 102 $\pm$ 3 & 1.0 $\pm$ 0.1 & 1.28 $\pm$ 0.07 & \\
15 & 9.2 & 68 $\pm$ 4 & 1.7 $\pm$ 0.1 & 1.03 $\pm$ 0.03 & \\
16 & 10.9 & 69 $\pm$ 4 & 2.0 $\pm$ 0.1 & 1.02 $\pm$ 0.02 & \\
17 & 9.9 & 65 $\pm$ 3 & 1.7 $\pm$ 0.1 & 1.33 $\pm$ 0.12 & \\
18 & 12.1 & 61 $\pm$ 5 & 1.2 $\pm$ 0.1 & 1.11 $\pm$ 0.05 & \\
19 & 7.9 & 80 $\pm$ 4 & 1.2 $\pm$ 0.2 & 1.41 $\pm$ 0.38 & \\
20 & 10.4 & 51 $\pm$ 2 & 1.0 $\pm$ 0.3 & 1.18 $\pm$ 0.06 & \\
21 & 12.7 & 74 $\pm$ 2 & 1.8 $\pm$ 0.1 & 1.33 $\pm$ 0.08 & \\
22 & 13.6 & 155 $\pm$ 7 & 0.2 $\pm$ 0.1 & 0.65 $\pm$ 0.04 & \\
\hline
\hline
\end{tabular}
\\[2mm] 
\begin{flushleft}
{\bf Note.} 
(1) KURVS ID;
(2) Maximal extent of the observed rotation curve;
(3) Intrinsic velocity dispersion, corrected for instrumental broadening and beam smearing (see text for details);
(4) Rotational support;
(5) Outer rotation curve slope.
$^{\star}$ {While this galaxy displays the typical kinematics signatures of a rotating disc in KURVS observations, the {\it HST} imaging suggests that these are associated with the orbital motion of a merging pair (see Appendix \ref{A:KURVS data} for details). 
We therefore exclude this galaxy from the dark matter fraction analysis presented in Sections \ref{Sect:Inner_RCs} and \ref{Sect:fDM}.}
 
\end{flushleft}
\end{table}

\subsection{Outer rotation curve shapes}
\label{Sect:Outer_RCs}

\begin{figure*}
\begin{center}
\centering
\includegraphics[scale=0.28]{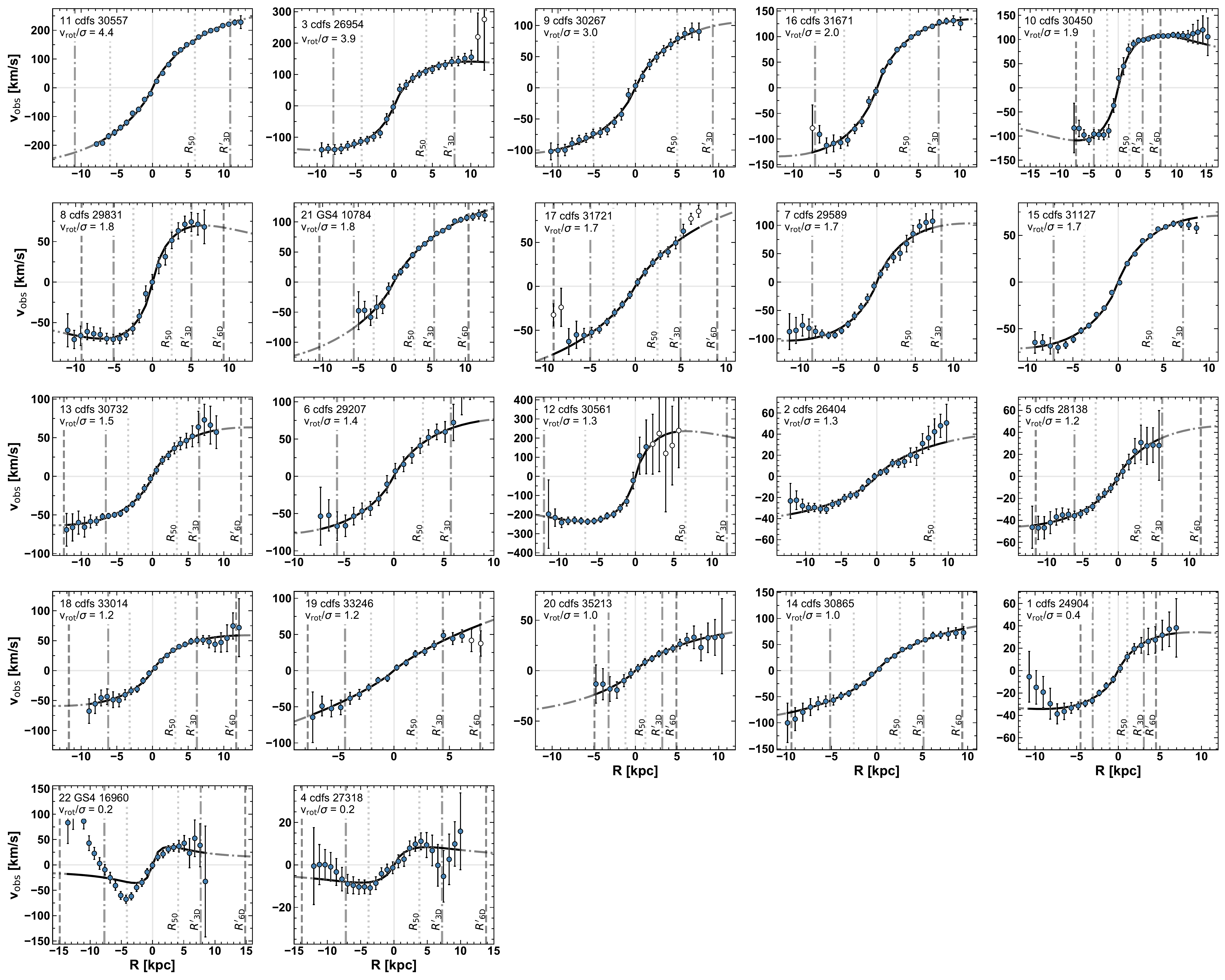}
\caption{Rotation curves and best-fitting 1D models for galaxies in the KURVS-CDFS sample. The sources are sorted in order of decreasing rotational support, quantified by the $v_{\rm rot}/\sigma$ ratio (see legend).
The white circles highlight noisy or unreliable data-points which have been discarded when fitting the 1D kinematic model after visual inspection of the cubes and PV-diagrams. These are either contaminated by sky lines or sample broad components that do not trace the overall kinematics of the disc.
The continuous black curve highlights the extent of the rotation curve that is probed by the data while the dash-dotted curve corresponds to the extrapolated model.
The dotted, dash-dotted and dashed vertical lines respectively mark the half-light radius measured from {\it HST} observations ($R_{\rm 50}$), and the seeing-convolved $3\times$ ($R'_{\rm 3D}$) and $6\times$ disc scale radii ($R'_{\rm 6D}$). Most galaxies have flat or rising rotation curves within $R'_{\rm 6D}$.}
\label{fig:all_rc}
\end{center}
\end{figure*}

We begin investigating the large-scale dynamics of $z \sim 1.5$ star-forming galaxies by studying the slopes of their rotation curves at large radii.
We show in Figure \ref{fig:all_rc} the rotation curves and best-fitting 1D models of KURVS-CDFS galaxies. 
Figure \ref{fig:all_rc} shows that most rotation curves are flat or continue to rise up to $\approx R'_{\rm 6D}$.

To perform a quantitative comparison with simulations, we measure the outer rotation curve shape with the parameter $t = v_{\rm R'_{\rm 6D}}/v_{\rm R'_{\rm 3D}}$ , following \citet{Tiley19}.
This corresponds to the ratio of the rotation velocity at $R'_{\rm 6D}$ and that measured at $R'_{\rm 3D}$ ($\approx 13$ kpc and $\approx 7$ kpc on average in our sample, see Section \ref{Sec:kin_pars} and Figure \ref{fig:all_rc}), roughly corresponding to the rotation velocity within and beyond the total rotation curve maximum, respectively. 

\begin{figure*}
\begin{center}
\centering
\includegraphics[scale=0.58]{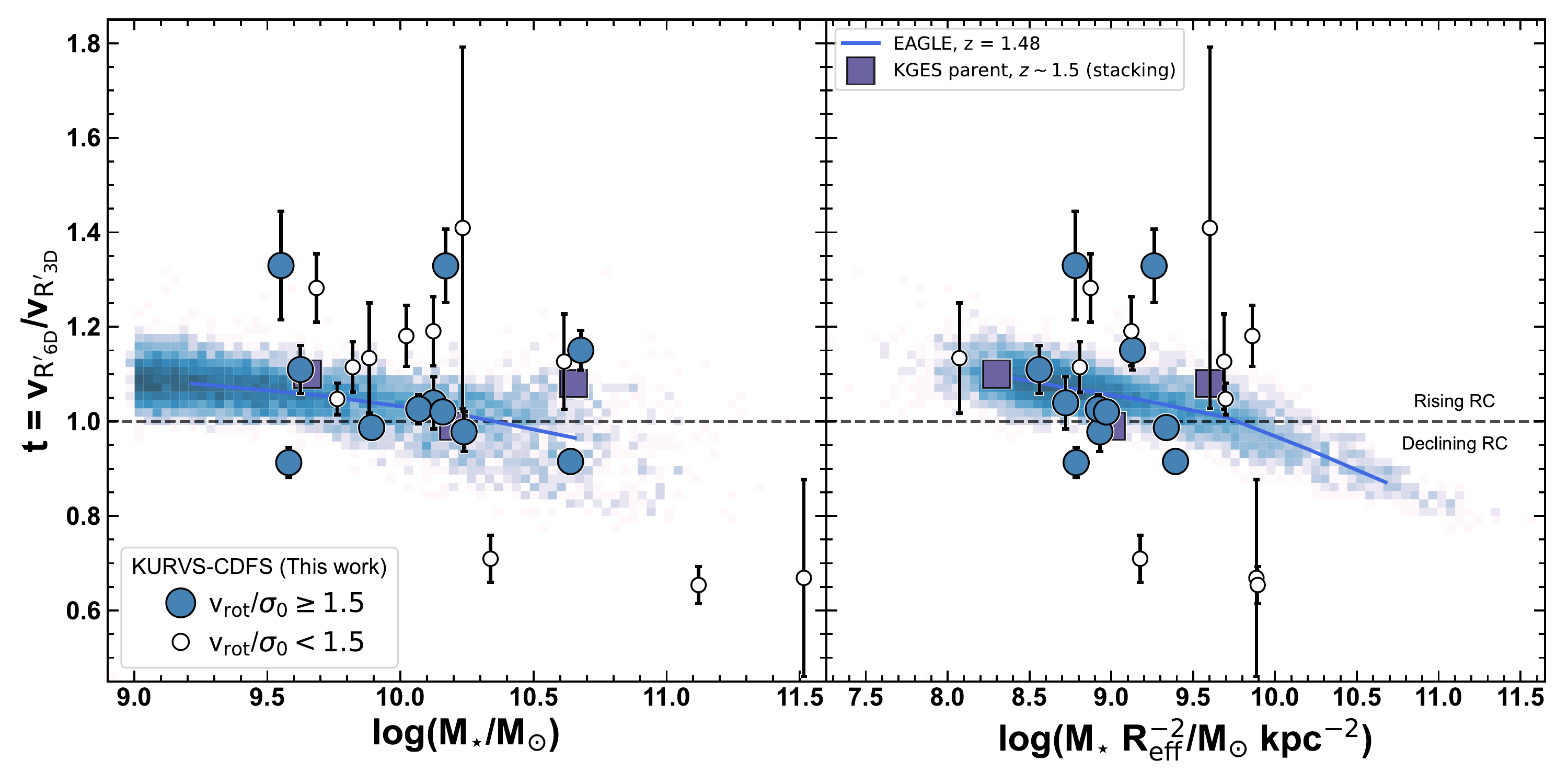}
\caption{Outer rotation curve slope, $t = v_{\rm R'_{6D}}/v_{\rm R'_{3D}}$, as a function of stellar mass (left) and stellar mass surface density (right). 
Measurements for simulated EAGLE galaxies at $z = 1.48$ are displayed with a 2D density histogram in blue. 
The running median for EAGLE galaxies is shown as a blue solid line. 
Measurements from stacked rotation curves at $z \sim 1.5$ from the KGES survey are shown with purple squares for comparison with our previous work \citep{Tiley19}.  
Large filled blue circles and small white circles represent measurements for rotationally-supported and dispersion-dominated KURVS galaxies, respectively.
The majority of observed rotation curves are flat or rising out to $R'_{\rm 6D}$, suggesting that $z \sim 1.5$ star-forming galaxies with stellar masses log$(M_{\star}/M_{\odot}) \sim 10.2$ contain a large fraction of their total mass in the form of dark matter within this radius.
Only three objects show substantially declining rotation curves. However, these have perturbed velocity fields and are dispersion-dominated. Therefore, measurements of the kinematics for such galaxies are highly uncertain. 
The observed rotation curve slopes are overall consistent with measurements of simulated EAGLE galaxies at similar redshifts.}
\label{fig:t_param}
\end{center}
\end{figure*}

A flat rotation curve would correspond to $t = 1$, whereas a rising or falling curve would be indicated by $t > 1$ and $t < 1$, respectively.
Figure \ref{fig:t_param} shows the $t$ parameter for observed and simulated $z \sim 1.5$ galaxies as a function of stellar mass and stellar mass surface density.
For consistency with the method used for observed galaxies, here we compute $t$ parameter on individual galaxies in the EAGLE simulation, selected to have log($M_{\star}/\mathrm{[M_{\odot}]}) \geqslant 9$ at $z = 1.48$ (\citealt{Schaller15}, see section 5.2.1 in \citealt{Tiley19} for how these circular velocity curves have been extracted).
As already discussed by \cite{Tiley19}, most galaxies in EAGLE have stellar masses log($M_{\star}/\mathrm{M_{\odot}}) \lesssim 10.3$ and display flat or rising rotation curves out to $R'_{\rm 6D}$. 
At the same time, the $t$ parameter shows a weak correlation and increased scatter with stellar mass (left panel in Figure~\ref{fig:t_param}), such that a larger fraction of massive galaxies in EAGLE have declining rotation curves at large radii. 
The correlation appears to be stronger with the stellar mass surface density (right panel in Figure \ref{fig:t_param}) and with a tighter scatter, such that most EAGLE galaxies with log($M_{\star}/R^2_{\rm eff}/\mathrm{[M_{\odot}/{\rm kpc}^2]}) \gtrsim 10$ display declining rotation curves out to $R'_{\rm 6D}$. 
EAGLE galaxies with significantly declining rotation curves ($t \leqslant 0.95$) are massive (log($M_{\star}/\mathrm{[M_{\odot}]}) \gtrsim 10$) and compact ($R_{\rm eff}  \lesssim 1.3$ kpc). These are a small fraction of the full EAGLE sample considered here (6~per~cent), but represent the 25~per~cent of simulated galaxies beyond log($M_{\star}/\mathrm{[M_{\odot}]}) \gtrsim 10$.
These objects have declining rotation curves at these spatial scales because this is also the scale where the galaxy transitions from being locally baryon-dominated to dark matter-dominated. 
These simulated galaxies present flat circular velocity curves beyond $R'_{\rm 6D}$.
We stress here that for the simulated galaxies we use circular velocity curves, and these are not influenced by perturbed gas kinematics. This explains the tight scatter of EAGLE galaxies in Figure \ref{fig:t_param}.

Focusing on the properties of observed galaxies, three KURVS galaxies have significantly declining rotation curves at large scales. 
Of these, one is a massive galaxy with $v_{\rm rot}/\sigma_0 \sim 1.3$ and a prominent bulge, as suggested by its {\it HST} imaging and its high Sersic index ($n_{\rm Ser} = 2.5$).
The integrated spectrum and kinematic map of this galaxy both indicate the presence of a prominent broad component ($\Delta v \gtrsim 300$ km s$^{-1}$), possibly tracing outflowing gas.
This suggests that an outflow or non-circular motions might be perturbing the kinematics of this object. 
The other two objects have $v_{\rm rot}/\sigma_0 \sim 0.2$ and a highly-perturbed velocity field (see Appendix~\ref{A:KURVS data}).
Furthermore, one of these galaxies is detected in X-rays and has an intrinsic 0.5 - 0.7 keV luminosity of $L_{\rm X-ray} \approx 1.1 \times 10^{42} \ \mathrm{erg s}^{-1}$ in the \cite{Luo17} catalogue.
This galaxy is also detected in the far-infrared and has a total far-infrared luminosity $L_{\rm IR} \approx 3 \times 10^{12} \mathrm{L}_{\odot}$  \citep{Birkin23}.  
This suggests that AGN and/or a merger-driven starburst are powering the H$\alpha$ emission of this object and its kinematics are highly perturbed.

Measurements of $t$ for KURVS-CDFS galaxies are reported in Table \ref{tab2:kinematics}.
The galaxies in our sample display overall rising rotation curve profiles at large radii, in broad agreement with the trend predicted from simulations and previous results from stacking \citep{Tiley19}. 
Our sample does not show any clear trend of declining rotation curves as a function of stellar mass or stellar mass surface density. 
This is, however, expected, because our galaxies probe a relatively narrow stellar mass/stellar mass surface density regime around log$(M_{\star}/\mathrm{M_{\odot}}) \sim 10.2$ and log($M_{\star}/R^2_{\rm eff}/\mathrm{M_{\odot}/kpc^2}) \sim 9$ as a result of our selection (see Section \ref{Sect:Selection}).
Fourteen galaxies have $t \sim 1$ within the 1$\sigma$ uncertainty, and hence flat rotation curves within $R'_{\rm 6D}$, corresponding to 64~per~cent of the KURVS-CDFS sample. 
Four sources or 14~per~cent of the sample have significantly rising rotation curves ($t \geqslant 1.2$) and are outliers from the simulated galaxies' trend at $\gtrsim 1 \sigma$. 
This does not seem to be associated with incorrect measurements of the disc scale radius, as this quantity is consistent with the mass-size relation in all but one of these galaxies.
Furthermore, high-resolution imaging from {\it HST} does not show any indication of extended faint components, suggesting that current measurements of the disc scale radius are robust. 
Future deep observations from {\it JWST} will allow us to improve measurements of the baryonic disc scale radius.
Four objects show declining rotation curves and $t < 1$ at $\gtrsim 2 \sigma$ significance, corresponding to the 14~per~cent of the sample. 
Three of these objects show substantially declining rotation curves, having $t \sim 0.7$, at $\gtrsim 5\sigma$ significance. 
These sources have dispersion-dominated dynamics and perturbed velocity fields.
We therefore conclude that there is an overall good agreement between the outer rotation curve shapes of observed and simulated $z \sim 1.5$ galaxies at log$(M_{\star}/\mathrm{M_{\odot}}) \sim 10.2$.

{We note that 12 galaxies have a rotation curve that extends up or beyond $R'_{\rm 6D}$, when accounting for uncertainties in this quantity. To test the robustness of our results against extrapolations of the rotation curve at large radii in the remaining of the sample, we measure the ratio $v_{\rm R'_{\rm 6D}}/v_{\rm H\alpha, max}$, where $v_{\rm H\alpha, max}$ represents the velocity at the maximal extent of the rotation curve. We measure a median $v_{R'_{\rm 6D}}/v_{\rm H\alpha, max} = 1.01^{+0.09}_{-0.07}$ where the uncertainties represent the $16^{\rm th}$ to $84^{\rm th}$ interquartile range. This suggests a marginal effect of extrapolations on measurements of the outer rotation curve slope discussed in this section.
Measurements of the rotation velocity at the relevant radii are indicated in Appendix \ref{A:KURVS data}.
To further quantify the impact of extrapolations on our conclusions, we measure the outer rotation curve slope as the ratio between the rotation velocity around $R'_{\rm 4D}$, which is reached by the majority of galaxies in the KURVS-CDFS sample. Measuring the outer rotation curve slope around $R'_{\rm 4D}$ does not affect our conclusions, suggesting that our results are robust against extrapolations of the rotation curve at large radii. }

\begin{figure}
\begin{center}
\centering
\includegraphics[scale=0.4]{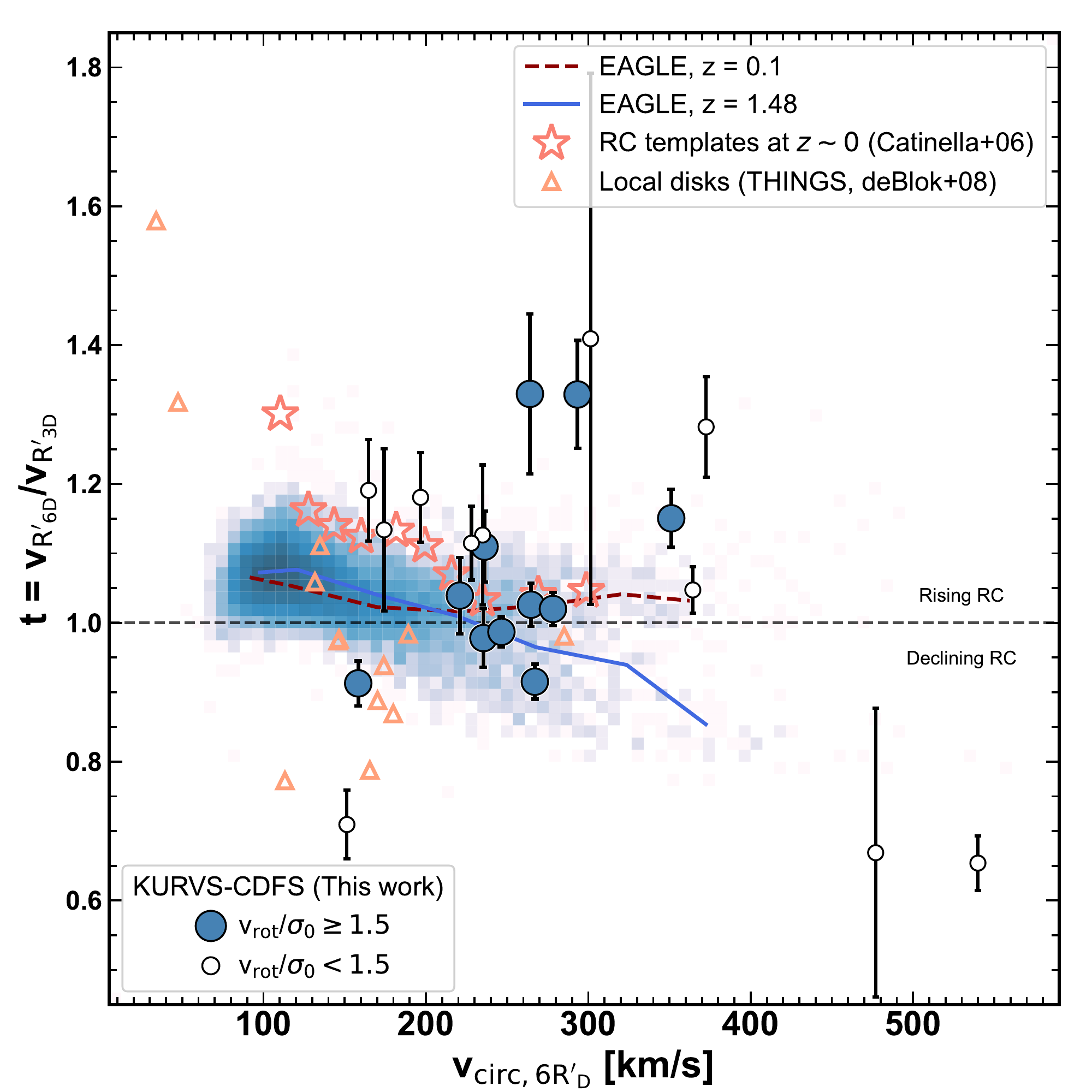}
\caption{ Outer rotation curve slope as a function of circular velocity at $R_{\rm 6D}$ for local and high-redshift star-forming galaxies.
The 2D density histogram shows measurements for simulated EAGLE galaxies at $z = 0.1$ and $z = 1.48$, with the coloured lines highlighting the running median at each redshift as specified in the legend.
The colour code for KURVS-CDFS galaxies is analogous to Figure \ref{fig:t_param}.
Open light-orange triangles show measurements for local discs from the THINGS survey \citep{deBlok08, Trachternach08}, while the open pink stars indicate the rotation curve templates of local star-forming discs from \citet{Catinella06}.
Outer rotation curve slopes of $z \sim 1.5$ star-forming galaxies are similar to measurements of local discs with similar rotational velocities suggesting little evolution with cosmic time at fixed dynamical mass.}
\label{fig:t_param_z0}
\end{center}
\end{figure}

We compare our measurements with the outer rotation curve shapes of local star-forming discs in Figure \ref{fig:t_param_z0}, where we show the $t$~parameter as a function of the circular velocity for $z \sim 1.5$ discs and local disc galaxies \citep{deBlok08, Trachternach08, Catinella06}.
For details on measurements of the circular velocity and $t$ in local galaxies see Appendix \ref{A2:RCs_z0}.
This figure shows that local and high-redshift star-forming galaxies in a similar circular velocity regime have diverse shapes, with mostly increasing or flat rotation curves at large radii. 
This was already seen in the \citet{Tiley19} stacking analysis.
This plot suggests little redshift evolution of the large-scale kinematic properties of galaxies at fixed circular velocity, similarly to what is found in simulated EAGLE galaxies and in previous studies   \citep[e.g.][]{Harrison17}.

\subsection{Inner kinematics}
\label{Sect:Inner_RCs}

We now compute the dark matter fraction within the effective radius, $f_{\rm DM}(\leqslant R_{\rm eff})$, which provides indications on the processes regulating the mass assembly of galaxies as well as the interplay between baryons and the surrounding dark matter halo.
We restrict this analysis to the eleven rotationally-supported ($v_{\rm rot} / \sigma_{0} \geqslant 1.5$) KURVS-CDFS galaxies, to minimise the contribution of turbulent motion to the galaxy dynamics.
Furthermore, we exclude from this analysis an interacting system of two galaxies which is unresolved with the seeing of our observations. 

To measure the dark matter fraction within the effective radius we need to subtract the contribution of baryons to the rotation curve at this radial scale. 
However, the effective radii of KURVS-CDFS galaxies are $\sim 1.5$ times the size of the KMOS PSF, on average.
This implies that, while the outer rotation curve slopes can be robustly recovered with minimal corrections and minimal modelling, the inner rotation curve shapes are heavily impacted by beam smearing, preventing an accurate kinematic decomposition at these small spatial scales.
To correct the inner rotation curve for the effect of beam smearing we use the \textsc{galpak$^{\rm 3D}$} tool \citep{Bouche15}, which allows us to obtain an ``intrinsic'' velocity field (i.e., corrected for the PSF and line spread function, LSF) by fitting a parametric model to a data cube. 
We adopt an exponential disc light profile with a Gaussian thickness of scale height $0.15 R_{\rm eff}$. 
To model the kinematics, we use a Freeman exponential disc for consistency with our analysis of the outer disc dynamics.
We fit this model to the continuum-subtracted KMOS cubes.
We extract rotation curves and velocity dispersion profiles from the PSF- and LSF-convolved best-fitting \textsc{galpak$^{\rm 3D}$} velocity and velocity dispersion maps, respectively, by applying the same method that was described in Section \ref{Sec:kin_pars}.
We accept the convolved rotation curve and velocity dispersion profile that minimise the weighted $\chi^2$ with respect to the observed profiles as the best-fitting solution.
This is because we want our best-fitting solution to be driven by the dynamics of the galaxy, which can be accurately reproduced with a smooth rotating disc, as opposed to the flux distribution which is clumpy and hence difficult to model with a smooth light profile.
We finally extract a ``beam smearing-corrected'' rotation curve from the deconvolved velocity map associated with our best-fitting solution, which we consider in the rest of the analysis presented in this section.
{We stress that, while forward-modelling techniques allow us to correct for the effects of beam smearing to recover the inner rotation curve shape, these models involve a large number of free parameters which can result in covariance between the best-fitting model parameters. 
The degeneracy and level of covariance depends on the intrinsic extent of the galaxy with respect to the seeing, its intrinsic brightness, and its H$\alpha$ morphology and emission line profile. 
However, while fixing the inclination to the HST value does not affect our results, we find strong covariance between the effective radius and the inclination in most cases, and this reflects the seeing-limited nature of KURVS observations. 
This is expected, given that the covariance between parameters is data- and seeing-specific \citep{Bouche15}.}
We therefore decide to adopt results from \textsc{galpak$^{\rm 3D}$} only to model the inner regions of the galaxies' discs.

Section \ref{Subsec:sigma_0} shows that turbulent motions provide a significant contribution to the dynamics of KURVS-CDFS galaxies, even when considering sources with $v_{\rm rot}/\sigma_0 \geqslant 1.5$ as in this section. 
As a result, turbulent motions partly compensate the gravitational force, and the rotation velocity is less than the circular velocity. 
Hence, we must correct the ``beam smearing corrected'' rotation curves for pressure support. 
We derive circular velocities using the \citet{Burkert10} formalism, considering as our characteristic velocity dispersion $\sigma_{0}$ the value from the H$\alpha$ velocity dispersion profiles (see Section \ref{Sec:kin_pars} and Table \ref{tab2:kinematics}).
Here we use observational estimates of the velocity dispersion, rather than the value output by \textsc{galpak$^{\rm 3D}$}, because this model does not allow us to fully capture the intrinsic complexity of the clumpy H$\alpha$ flux distribution, and hence of the turbulence profile, in high-redshift discs. 
Future, high-resolution observations with adaptive-optics equipped integral field spectrographs such as ERIS or high spatial resolution observations with ALMA will allow us to improve our estimates of the velocity dispersion against beam smearing effects, and minimise this source of scatter in measurements of the dark matter fraction. 

Finally, we model the baryonic contribution to this ``intrinsic'' (i.e., corrected for the PSF, LSF, and pressure support) rotation curve with a Freeman disc with mass corresponding to the sum of the stellar mass and a 40\% molecular gas fraction\footnote{Here we define the molecular gas fraction as $f_{\rm mol} = M_{\rm mol} / (M_{\star} + M_{\rm mol})$}, which is the typical value expected from scaling relations at the average stellar mass and redshift of our sample \citep{Tacconi20}.
{We tested however that evaluating the molecular gas fraction from the \cite{Tacconi20} scaling relation at the stellar mass of each galaxy does not affect the results on measurements of the dark matter fraction within the effective radius.}
We assume that the baryonic component has an effective radius corresponding to that measured on {\it HST} imaging in the near-infrared.
We adopt a Freeman thin disc to model the baryonic contribution to the rotation curve for simplicity.
In reality, high-redshift galaxies predominantly have turbulent and clumpy discs, but more observations are needed to constrain the discs' intrinsic thickness.
While the shape of the rotation curve for a thick disc is similar to that of the thin disc approximation at large radii, a disc model with intrinsic thickness $q_0 = 0.2$ would decrease the velocity of the baryonic component by $\sim 10$~per~cent at the effective radius \citep[e.g.][]{Casertano83}. This is within the uncertainties for our estimated dark matter fractions at these scales, and does not affect our conclusions. A thick disc would also modify the shape of the baryonic rotation curve at radial disc scales, hence accurate modelling of this component is required to derive dark matter profiles of high-redshift galaxies. This is beyond the scope of this work, and we defer measurements of the dark matter profiles in the full KURVS sample to a future paper (Dudzevi\v{c}i\={u}t\.{e} et al., in prep).

\subsection{Dark matter fraction within the effective radius}
\label{Sect:fDM}
We use the beam-smearing and pressure-support corrected rotation curves to compute the dark matter fractions at the effective radius as:
\begin{equation}
f_{\rm DM}(\leqslant R_{\rm eff}) = 1 - \frac{v_{\rm bar}^2(R_{\rm eff})}{v_{\rm tot}^2(R_{\rm eff})}.
\label{fdm}
\end{equation}
Here $v_{\rm tot}$ is the velocity of the pressure-support and inclination-corrected best-fit rotation curve from \textsc{galpak$^{\rm 3D}$}, and $v_{\rm bar}$ is the baryon contribution to the rotation curve as described in the previous section.
Measurements of the dark matter fraction for rotationally-supported KURVS-CDFS galaxies are reported in Table \ref{tab3:fdm}. 
Uncertainties quoted here correspond to the standard deviation of the dark matter fraction measured by varying the total and baryonic rotation curves within the 1$\sigma$ errors 1000 times and do not consider systematics variations in the baryon mass. Varying the baryon mass by $\pm$0.2 dex would imply a $\mp$ variation in the dark matter fraction, on average.

\begin{figure*}
\begin{center}
\centering
\includegraphics[scale=0.58]{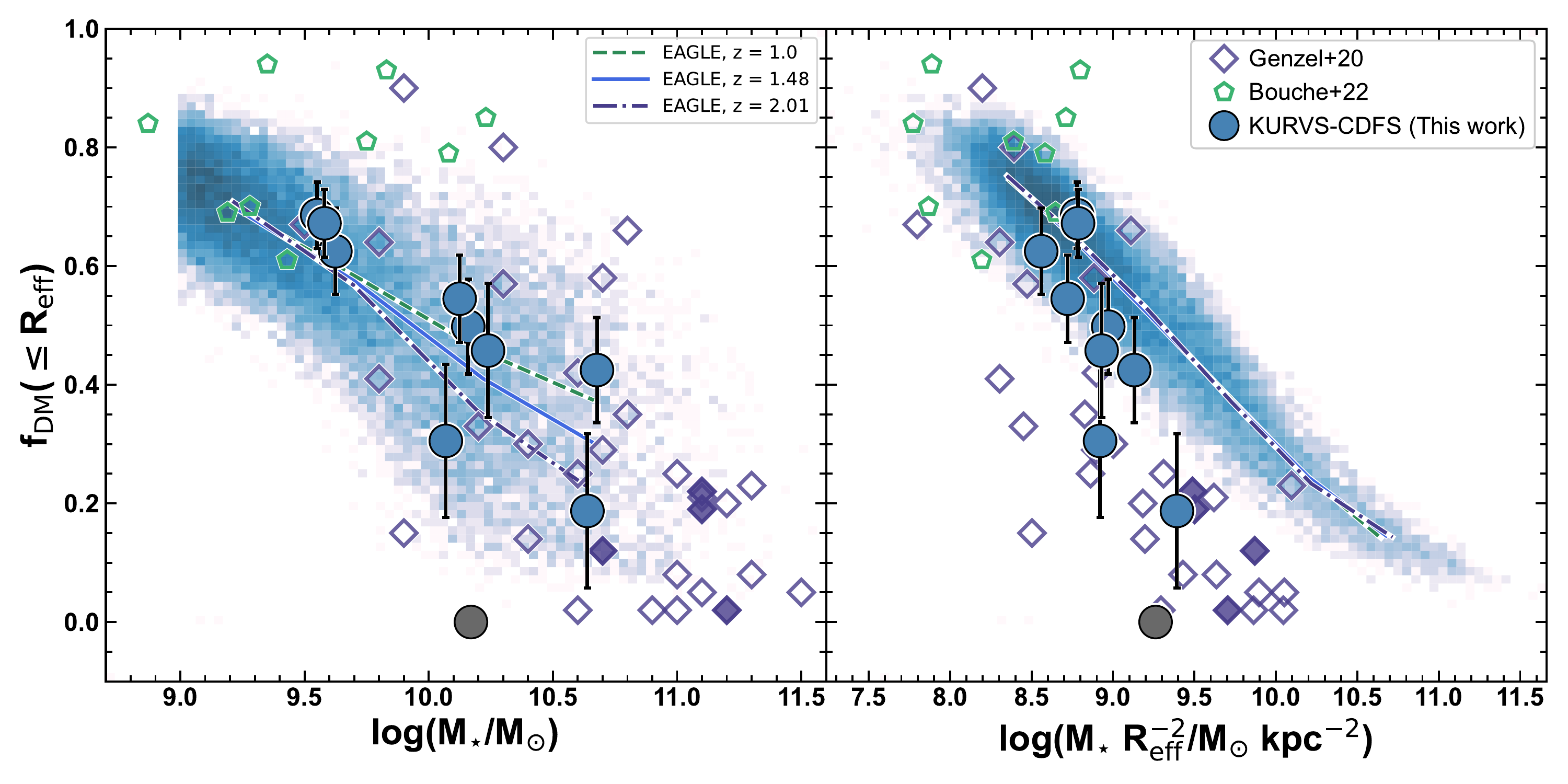}
\caption{Dark matter fraction within the disc effective radius for high-redshift star-forming galaxies as a function of stellar mass (left) and stellar mass surface density (right). 
The 2D density histogram shows measurements for simulated EAGLE galaxies at $z$ = 1.0, 1.48 and 2.01. The coloured lines show the running median at each redshift in EAGLE.
The open green pentagons show the \citet{Bouche22} sample of $z \sim 0.9$ star-forming galaxies. 
The open purple diamonds indicate the \citet{Genzel20} sample at $0.67 \lesssim z \lesssim 2.45$, with filled symbols highlighting the sources that are in the same redshift range as our sample.
Large filled circles represent rotationally-supported galaxies in KURVS-CDFS, with the grey data point indicating a source whose kinematics are not well-described by \textsc{galpak}$^{\rm 3D}$. 
Simulated EAGLE galaxies show a trend of decreasing dark matter fraction with increasing stellar mass, albeit with large scatter. Observed galaxies follow the same trend.
There is a much tighter anti-correlation between the dark matter fraction and the stellar mass surface density in EAGLE. 
Observations follow a similar trend up to log$(\Sigma_{\star}/M_{\odot}/{\rm kpc}^{-2}) \sim 9.2$. 
At higher stellar mass surface densities, however, EAGLE overpredicts the dark matter fraction of observed galaxies by a factor of $\sim 3$. }
\label{fig:fDM_Re_baryons}
\end{center}
\end{figure*}

The left panel of Figure \ref{fig:fDM_Re_baryons} shows the dark matter fraction at the effective radius as a function of stellar mass.
This figure shows that $f_{\rm DM}(R_{\rm eff}) \gtrsim 40 \%$ in most galaxies in our sample, while two galaxies show a lower dark matter fraction. 
Finally, one galaxy, KURVS-21, formally has zero dark matter fraction at the effective radius. 
This object is a compact and massive galaxy with a prominent bulge and an asymmetric rotation curve which is difficult to reproduce accurately with \textsc{galpak$^{\rm 3D}$}. As a result, the dark matter fraction for this object is uncertain and we flag the measurement in the plot.
Therefore, we conclude that the dark matter provides a significant contribution to the dynamics of rotationally-supported KURVS-CDFS galaxies already at scales of the effective radius.
The dark matter fraction of our sample is intermediate between that of low-mass star-forming galaxies at $z \sim 0.9$ \citep[][open green pentagons in Figure \ref{fig:fDM_Re_baryons}]{Bouche22} and massive star-forming galaxies at $0.6 \leqslant z \leqslant 2.4$ \citep[][violet diamonds in Figure \ref{fig:fDM_Re_baryons}]{Genzel20}.
Taken together, this suggests that the dark matter fraction at $R_{\rm eff}$ decreases with increasing stellar mass, as already suggested in the literature \citep{Tiley19, Genzel17, Genzel20, Bouche22}. 
This observed trend appears to be in qualitative agreement with measurements in the EAGLE simulation, which also indicate an increased scatter at stellar masses larger than log$(M_{\star}/M_{\odot}) \sim 10.3$. 

The right panel of Figure \ref{fig:fDM_Re_baryons} shows the dark matter fraction within the effective radius as a function of stellar mass surface density.
{As already discussed in previous studies \citep[][see also \citealt{Wuyts16}]{Genzel20, Bouche22}, this correlation is much tighter than the one with the stellar mass. }
The observed data agree with measurements of simulated galaxies up to log($M_{\star} R^{-2}_{\rm eff}/ M_{\odot} kpc^{-2}) \sim 9.2$.
At higher stellar mass surface density, the EAGLE simulation over-predicts the dark matter content of galaxies by a factor of $\sim 3-4$.
The low dark matter fractions measured in high-redshift galaxies with high stellar mass and stellar mass surface density have been interpreted as an indication that their inner dark matter profiles deviate from the cusps predicted in the context of $\Lambda$CDM models \citep{Genzel20}, and other studies have also showed indications of cored dark matter profiles in lower mass galaxies in the distant Universe \citep{Bouche22, Sharma22}. We will explore this aspect in a future paper, using a detailed modelling of the baryonic component from spatially-resolved {\it HST} photometry (Dudzevi\v{c}i\={u}t\.{e} et al., in prep). 

\begin{table}
\centering
\caption{Dark matter fraction of rotationally-supported KURVS-CDFS galaxies.}
\label{tab3:fdm}
\begin{tabular}{cc} 
\hline
\hline
KURVS ID & $f_{\rm DM}(\leqslant R_{\rm eff})$  \\
(1) & (2) \\
\hline
11 & 0.42 $\pm$ 0.09 \\
13 & 0.62 $\pm$ 0.07 \\
15 & 0.31 $\pm$ 0.13 \\
16 & 0.50 $\pm$ 0.08 \\
17 & 0.69 $\pm$ 0.06 \\
21$^{\star}$ & 0.0  \\
3 & 0.19 $\pm$ 0.13 \\
7 & 0.46 $\pm$ 0.11 \\
8 & 0.67 $\pm$ 0.06 \\
9 & 0.54 $\pm$ 0.07 \\
\hline
\hline
\end{tabular}
\\[2mm] 
\begin{flushleft}
{\bf Note.} 
(1) KURVS-ID. 
(2) Dark matter fraction within the effective radius and $1 \sigma$ uncertainties. This is obtained on seeing- and pressure support- corrected rotation curves using equation~\ref{fdm}. Galaxies are sorted in order of decreasing $v_{\rm rot}/\sigma_0$. 
$^{\star}$Galaxy KURVS-21 has a highly asymmetric velocity field, that makes it difficult to find a best-fitting model with \textsc{galpak$^{\rm 3D}$}. 
While we report its dark matter fraction for completeness, this measurements is highly unreliable and we flag this galaxy in the relevant plots.
\end{flushleft}
\end{table}

\begin{figure}
\begin{center}
\centering
\includegraphics[scale=0.4]{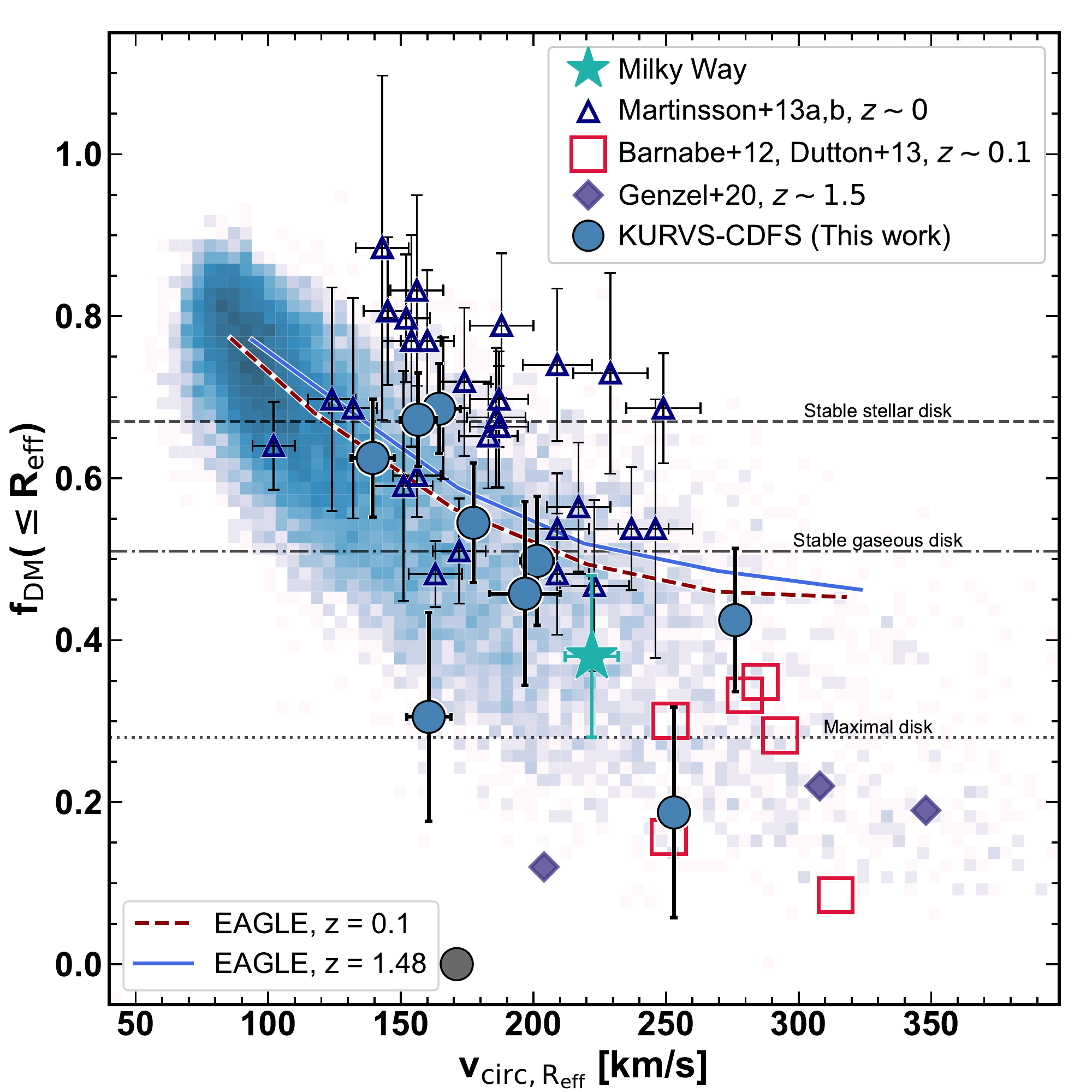}
\caption{ Dark matter fraction within the effective disc radius as a function of circular velocity at the effective radius for local and $z \sim 1.5$ star-forming galaxies.
The 2D density histogram and coloured lines show measurements for simulated EAGLE galaxies at $z$ = 0.1 and 1.48 as in Figure \ref{fig:t_param_z0}.
The open dark blue triangles show measurements of $z \sim 0$ spirals from \citet{Martinsson13a, Martinsson13b}, and the filled green star indicate the dark matter fraction of the Milky Way {\citep{BovyRix13, BlandHawthorn16}}.
The open red squares indicate measurements of massive spiral galaxies with prominent bulges at $z \sim 0.1$ from the SWELLS survey \citep{Barnabe12, Dutton13}.
The purple diamonds indicate the \citet{Genzel20} galaxies in the redshift range of our survey. 
Large filled circles represent rotationally-supported galaxies in KURVS-CDFS, with the grey data point indicating a galaxy whose kinematics are not well described by \textsc{galpak}$^{\rm 3D}$. 
The horizontal dotted line highlights the dark matter fraction of a maximal disc {(i.e., centrally dominated by baryons)} within $R_{\rm eff}$, $f_{\rm DM}(\leq {\rm R}_{\rm eff}) = 0.28 $.
{The horizontal dash-dotted and dashed grey lines show the dark matter fraction required for a gaseous and a stellar disc to be stable against axisymmetric instabilities \citep[$f_{\rm DM}(\leq {\rm R}_{\rm eff}) = 0.51 $ and $f_{\rm DM}(\leq {\rm R}_{\rm eff}) = 0.67 $ respectively, ][]{Efstathiou82}.}
The dark matter fraction of $z \sim 1.5$ discs is consistent with that of local discs in the same circular velocity interval. 
Therefore, this figure suggests that the dark matter content of galaxies is mostly associated with the galaxy stellar/dynamical mass with little or no trend with redshift.}

\label{fig:fDM_Re_vs_z0}
\end{center}
\end{figure}

Figure \ref{fig:fDM_Re_vs_z0} shows the dark matter fraction as a function of the circular velocity at $R_{\rm eff}$ for rotationally-supported star-forming galaxies at $z \sim 1.5$ and local discs. 
The dark matter content of $z \sim 1.5$ star-forming galaxies is similar to that of local galaxies with similar circular velocity, and it decreases as a function of this quantity. This trend seems to have little or no dependence on cosmic time.

{To quantify the difference in the dark matter fraction within the effective radius between $z \sim 1.5$ and local discs, we divide the  samples in two circular velocity bins and measure the average dark matter fractions in both populations. 
We find that $f_{\rm DM}(\leqslant R_{\rm eff}) = 0.57 \pm 0.16$ in KURVS galaxies with $v_{\rm circ} = 155 \pm 10$ km s$^{-1}$, and $f_{\rm DM}(\leqslant R_{\rm eff}) = 0.71 \pm 0.11$ in local discs from the \cite{Martinsson13a,Martinsson13b} sample having $v_{\rm circ} = 146 \pm 20$ km s$^{-1}$. 
On the other hand, we measure $f_{\rm DM}(\leqslant R_{\rm eff}) = 0.42 \pm 0.12$ in KURVS galaxies with $v_{\rm circ} = 221 \pm 40$ km s$^{-1}$, and $f_{\rm DM}(\leqslant R_{\rm eff}) = 0.57 \pm 0.23$ in local discs from the \cite{Martinsson13a,Martinsson13b} and \cite{Barnabe12, Dutton13} samples with $v_{\rm circ} = 227 \pm 40$ km s$^{-1}$. Here the $1 \sigma$ uncertainties indicate the standard deviation of the samples' distribution.
That is, our measurements indicate that rotationally-supported star-forming galaxies at $z \sim 1.5$ have $\sim 14 \% - 15\%$ lower dark matter fraction with respect to local discs with comparable circular velocities. However, this difference is less than $1 \sigma$ significant and the average dark matter fractions are consistent within the $1 \sigma$ uncertainties. 
We therefore conclude that there are no significant differences in the dark matter content of star-forming galaxies as a function of cosmic time, and that larger samples with ultra-deep observations at $z \sim 1.5$ are required for investigating systematic differences in this quantity as a function of redshift.}

Our observational results are consistent with measurements of simulated EAGLE galaxies, which show the same trend of decreasing dark matter fraction  as a function of the circular velocity and only a small evolution of the normalisation as a function of redshift. 

\section{Discussion and conclusions}
\label{Sect:Discussion}

In this work we use ultra-deep observations from the KURVS survey to characterise the dynamical properties of star-forming galaxies at $z \sim 1.5$.
The unprecedented depth of KURVS observations allows us measure robust rotation curve shapes out to $\sim 10-15$ kpc, which is near or beyond 6 times the disc scale radius.
Our analysis shows that most of these rotation curves are flat or rising, suggesting that dark matter dominates the dynamics of $z \sim 1.5$ star-forming galaxies at large radii, similar to discs with the same circular velocities in the local Universe and simulations (see Figures \ref{fig:t_param} and \ref{fig:t_param_z0}).
We also find that, once beam smearing and pressure support are corrected for, dark matter provides a significant contribution to the dynamics of rotationally-supported galaxies in our sample already within the stellar disc, again in broad agreement with measurements from cosmological simulations and in the local Universe (Figures \ref{fig:fDM_Re_baryons} and \ref{fig:fDM_Re_vs_z0}). 
We begin this section by discussing some caveats of measurements of the dark matter fraction in the distant Universe. 
After that, we discuss our results in the context of literature results and simulations.

\subsection{Caveats on measuring dark matter fractions in distant galaxies}

Ultra-deep spatially-resolved observations sampling the outer regions of the disc are an essential ingredient to characterise the outer rotation curve slope and dynamical properties of star-forming galaxies at the peak epoch of cosmic mass assembly.
These observations are critical to probe the spatial scales where dark matter is expected to provide the dominant contribution to the galaxies' dynamics, and hence measure the dark matter halo mass with small extrapolations and minimal degeneracies due to the baryonic disc.
At the same time, additional observational constraints on the kinematics and baryonic properties of galaxies are required to characterise the masses and density profiles of dark matter haloes in the distant Universe. 

High-redshift galaxies have a highly turbulent interstellar medium, as indicated by the increase in velocity dispersion with increasing redshift \citep[][see also right panel of Figure \ref{fig:Dynamics, vsigma_sigmaz}]{Turner17,Johnson18, Ubler19, Wisnioski19, Jimenez22}. 
This implies that measurements of the circular velocity in high-redshift discs are largely dependent on the pressure-support correction, especially for low-mass galaxies that have lower rotation velocities, and at large radii, because the correction increases as a function of radius \citep[e.g.,][]{Burkert10}.
Pressure support corrections for high-redshift galaxies assume a constant characteristic value for the velocity dispersion throughout the disc, and a vertical geometry which are currently poorly constrained by data (\citealt{Wellons20}, see also discussion in \citealt{Bouche22}). 
Adaptive optics assisted integral-field observations at high spatial resolution will be needed to improve our understanding of the turbulence and vertical structure of high-redshift discs, hence improving pressure support corrections for distant galaxies.
Additionally, observations of colder dynamical tracers such as the molecular gas will be important to understand the impact of feedback and outflows on measurements of the velocity dispersion, as these processes are expected to affect mostly the ionised gas component \citep[e.g.][]{Krumholz18, Girard21}.

The contribution of the molecular gas to the baryon mass budget is another important aspect to consider when measuring the dark matter content of distant galaxies.
Molecular gas fractions of high-redshift galaxies are substantial (perhaps up to $\sim60$~per~cent), and there are up to a factor of $\sim 6$ object-by-object variations \citep{Tacconi20}.
Furthermore, the molecular gas can be from $\sim 3$ times smaller to about twice as extended as the stellar disc \citep{Puglisi19}. 
Variations in the molecular gas content and extent can significantly impact measurements of dark matter fractions and profiles in distant star-forming galaxies.
These effects are particularly important for the most massive and centrally-concentrated galaxies, for which ALMA observations suggest 
$\sim 0.5-09$ dex lower gas fractions than expected from scaling relations, and compact molecular gas cores \citep{Tadaki17, Elbaz18, Franco20, Puglisi21b}. 
Moreover, other observational systematics, such as colour gradients reducing the stellar mass effective radius by up to a $\sim 25$~per~cent \citep{Suess22} or centrally-peaked dust attenuation profiles \citep[e.g.,][]{Tacchella18} can impact measurements of the dark matter halo properties at high redshift, especially for the most massive star-forming galaxies. 
Direct observations of the molecular gas and of the stellar continuum of high-redshift disks with ALMA and {\it JWST} are thus needed to quantify the baryonic contribution to observed rotation curves with high accuracy, and hence dark matter fractions \citep[e.g.,][]{Molina19}. 
Because these effects are expected to largely affect the most massive and centrally-concentrated galaxies, i.e. the objects showing the largest discrepancies from the predictions of models, obtaining direct constraints on the full baryon profile of these galaxies might help to alleviate the tension between models and observations.

Finally, beam smearing effects impact the study of the small-scale dynamics of distant galaxies, as extensively discussed in the literature \citep[e.g.][]{Cresci09,Bouche15,diTeodoro16}.
Here we use a parametric tool to correct inner rotation curves for beam smearing, following a similar procedure to that which has been broadly applied across various studies of high-redshift galaxies \citep[e.g. ][]{Sharma21,Genzel20,Price21,Rizzo21}. 
On the other hand, while parametric tools are an excellent resource for recovering the galaxy kinematics at small scales for seeing-limited observation, these tools have a large number of free parameters which imply large degeneracies in the results. 
Higher resolution observations will be key to link the outer and small-scale kinematics of distant galaxies and study their dark matter content within the optical radius with high accuracy.

In summary, ultra-deep rotation curves are the first, essential step in quantifying the dynamics of baryons and the radial distribution of dark matter in the distant Universe. 
However, a full sampling of the baryon mass and distribution at high spatial resolution is needed to quantify the properties of dark matter haloes at high redshift with high accuracy and quantify possible deviations between observations and theoretical models.

\subsection{Comparison with $z \sim 0$ discs and simulations}

Leaving aside the systematics discussed above, which are common to most observational estimates of the dark matter content of distant galaxies, we can start discussing our results in the context of other samples of local and high-redshift star-forming galaxies, as well as simulations.

Our study shows that the contribution of dark matter to the dynamics of galaxies decreases as a function of the stellar mass/stellar mass surface density, similarly to what is suggested by observations in the local Universe \citep{CourteauDutton15}.
In fact, Figures \ref{fig:t_param_z0} and \ref{fig:fDM_Re_vs_z0} show that the outer rotation curve shapes and dark matter content of galaxies seem to have little evolution with cosmic time at fixed circular velocity. 
These observational results are in agreement with measurements of simulated EAGLE galaxies, which indicate high dark matter fractions within the effective radius in galaxies with log$(M_{\star}/\mathrm{M_{\odot}}) \geqslant 9$ at nearly all redshifts up to $z \sim 2$.
This is not incompatible with literature results, showing that high-redshift discs have significantly lower dark matter fractions than predicted from cosmological models \citep{Genzel17, Genzel20, Price21}.
In fact, KURVS observations sample the scatter of the $z \sim 1.5$ main sequence and mass-size relation at log$(M_{\star}/\mathrm{M_{\odot}}) \sim 10 - 10.5$ (see Figure \ref{fig:Selection, MS_MSize}), which is close to the knee of the stellar mass function of the star-forming population in the same redshift range (log$(M_{\star}/M_{\odot}) \sim 10.42$; \citealt{Davidzon17}).
On the other hand, the \citet{Genzel20} sample in the same redshift range probed by KURVS observations have $M_{\star} \sim 10^{11.5} \ \mathrm{M_{\odot}}$ (see Figures \ref{fig:fDM_Re_baryons} and \ref{fig:fDM_Re_vs_z0}), and hence probe the high-mass end of the main sequence at $z \sim 1.5$.
The high-mass end of the main sequence is where ``typical discs'' become rarer, because this regime samples beyond the characteristic $M_{\star}$ of the stellar mass function, and where galaxies are expected to grow massive bulges and subsequently quench \citep{DekelBirnboim06, Zolotov15}.
Therefore, such massive galaxies might not be representative of the bulk of the star-forming population and the tension between observations and simulations might be related to quenching mechanisms and/or a poor understanding of bulge growth mechanisms and spheroid formation.
Indeed, \citet{Genzel20} noted that the preference for a cored dark matter profile is strongly correlated with the central bulge mass.

The idea that differences between observed and simulated dark matter fractions are associated with the baryonic properties of galaxies, rather than being a problem of the underlying dark matter backbone, is supported by the fact that these arise when considering the dependence of $f_{\rm DM}(R_{\rm eff})$ on the stellar mass surface density (right panel of Figure \ref{fig:fDM_Re_baryons}).
Indeed, cosmological models such as EAGLE struggle to reproduce the sizes and light profiles of massive galaxies at $z \sim 0$ \citep{RodriguezGomez19,deGraaff22}, and Illustris-TNG recovers the low dark matter fractions of massive star-forming disks at high-redshift, but for too-small effective radii \citep{Ubler21}.
It is plausible that the discrepancies are driven by the sub-grid modelling of physical processes of galaxy formation and feedback, and its complex interplay with the underlying dark matter halo distribution.
In fact, state-of-the-art simulations such as EAGLE and Illustris-TNG seem to over-predict the dark matter content of 
massive spirals in the local Universe, possibly as a result of the feedback implementation \citep[the `failed feedback problem'; ][]{Posti19, Marasco20}. 
Future work will thus also be required from the simulation perspective to understand what drives the difference between the dark matter halo and baryonic properties of simulated and observed galaxies in the distant Universe.

\section*{Acknowledgements}
The authors would like to thank the referee for their constructive report on our paper.
This work was supported by STFC through grants ST/T000244/1 and ST/P000541/1.
S.G. acknowledges the support of the Cosmic Dawn Center of Excellence funded by the Danish National Research Foundation under the grant 140.
L.C. acknowledges support from the Australian Research Council Discovery Project and Future Fellowship funding schemes (DP210100337, FT180100066).
C.H. acknowledges funding from a United Kingdom Research and Innovation grant (code: MR/V022830/1).
E.I. acknowledge funding by ANID FONDECYT Regular 1221846.
D.O. is a recipient of an Australian Research Council Future Fellowship (FT190100083) funded by the Australian Government.
K.A.O. acknowledges support by the European Research Council (ERC) through Advanced Investigator grant to C.S.~Frenk, DMIDAS (GA~786910).
A.P. and M.S. thank Ian Smail for useful discussions and valuable inputs on the analysis.
A.P. also thanks Nicolas Bouche for useful discussions regarding the \textsc{galpak}$^{\rm 3D}$ code.
This work has made use of NASA's Astrophysics Data System.

\section*{Data availability}
The data used in this paper are available through the ESO archive.

\bibliographystyle{mnras}
\bibliography{kurvs_bib} 


\appendix

\section{The KURVS-CDFS dataset}
\label{A:KURVS data}

We include in this section the kinematic maps (Figures \ref{figA1:kurvs_dataset2}-\ref{figA3:kurvs_dataset4}) and position-velocity diagrams (Figure \ref{fig:kurvs_pv}) for all galaxies in the KURVS-CDFS pointing analysed in this work.
{We provide measurements of the maximal rotation curve extent and velocity at the last observed data-point on the rotation curve, as well as measurements of the rotation velocity at $R'_{\rm 3D}$ and $R'_{\rm 6D}$ in Table \ref{tab3:radii}.}

\begin{table}
\centering
\caption{{Rotational velocities of KURVS-CDFS galaxies at different radii.}}
\label{tab3:radii}
\begin{tabular}{ccccc} 
\hline
\hline
KURVS ID & R$_{\rm H\alpha, max}$  & $v_{\rm H\alpha, max}$ & $v_{\rm R'_{\rm 3D}}$ & $v_{\rm R'_{\rm 6D}}$ \\
 & kpc & km s$^{-1}$  & km s$^{-1}$ & km s$^{-1}$   \\
  (1) & (2) & (3) & (4)  & (5) \\
\hline
  1    &    10.7 &    49.3   $\pm$  8.9  &    40.9   $\pm$  1.8   &     42.9   $\pm$  2.1         \\  
  2    &    12.1 &    48.4   $\pm$  13.0 &    52.4   $\pm$  7.6   &     59.4   $\pm$  14.1        \\  
  3    &    11.7 &    209.8  $\pm$  14.1 &    217.1  $\pm$  3.7   &     198.7  $\pm$  5.0         \\  
  4    &    12.0 &    8.0    $\pm$  2.0  &    10.4   $\pm$  1.0   &     7.4    $\pm$  1.1         \\  
  5    &    12.1 &    55.6   $\pm$  12.0 &    46.3   $\pm$  4.8   &     55.1   $\pm$  6.7         \\  
  6    &    8.1  &    82.1   $\pm$  36.4 &    77.2   $\pm$  6.8   &     87.0   $\pm$  7.3         \\  
  7    &    11.4 &    105.4  $\pm$  9.9  &    104.8  $\pm$  5.0   &     102.5  $\pm$  7.4         \\  
  8    &    11.0 &    68.8   $\pm$  6.7  &    79.4   $\pm$  16.7  &     72.5   $\pm$  15.5        \\  
  9    &    10.2 &    136.9  $\pm$  8.3  &    137.7  $\pm$  6.8   &     143.1  $\pm$  11.5        \\  
  10   &    15.2 &    93.4   $\pm$  16.8 &    115.9  $\pm$  4.4   &     114.4  $\pm$  4.0         \\  
  11   &    12.4 &    246.6  $\pm$  5.2  &    240.0  $\pm$  7.7   &     276.1  $\pm$  7.1         \\  
  12   &    11.2 &    258.2  $\pm$  78.0 &    256.6  $\pm$  78.6  &     171.7  $\pm$  53.3        \\  
  13   &    12.1 &    94.3   $\pm$  10.7 &    85.2   $\pm$  3.4   &     94.6   $\pm$  5.4         \\  
  14   &    9.8  &    105.6  $\pm$  7.3  &    80.8   $\pm$  7.6   &     103.6  $\pm$  9.9         \\  
  15   &    9.2  &    112.2  $\pm$  3.6  &    111.0  $\pm$  2.8   &     113.9  $\pm$  3.3         \\  
  16   &    10.9 &    139.4  $\pm$  6.1  &    136.0  $\pm$  3.0   &     138.7  $\pm$  1.2         \\  
  17   &    9.9  &    120.9  $\pm$  10.7 &    86.5   $\pm$  6.6   &     115.0  $\pm$  8.6         \\  
  18   &    12.1 &    76.1   $\pm$  8.6  &    68.1   $\pm$  2.7   &     75.9   $\pm$  4.5         \\  
  19   &    7.9  &    95.4   $\pm$  19.6 &    67.2   $\pm$  17.9  &     94.6   $\pm$  25.3        \\  
  20   &    10.4 &    80.2   $\pm$  28.5 &    44.9   $\pm$  11.7  &     53.0   $\pm$  13.8        \\  
  21   &    12.7 &    145.6  $\pm$  15.7 &    100.4  $\pm$  6.4   &     133.4  $\pm$  8.7         \\  
  22   &    13.6 &    28.9   $\pm$  3.2  &    42.1   $\pm$  2.9   &     27.5   $\pm$  1.8         \\  
\hline
\hline
\end{tabular}
\\[2mm] 
\begin{flushleft}
{\bf Note.} 
(1) KURVS ID;
(2) Maximal extent of the observed rotation curve;
(3) Inclination-corrected velocity at the maximal extent of the observed rotation curve;
(4) Inclination- and beam-smearing corrected velocity at $R'_{\rm 3D}$;
(5) Inclination-corrected velocity at $R'_{\rm 6D}$;
\end{flushleft}
\end{table}

\begin{figure*}
\includegraphics[angle=90,origin=c,scale=0.45]{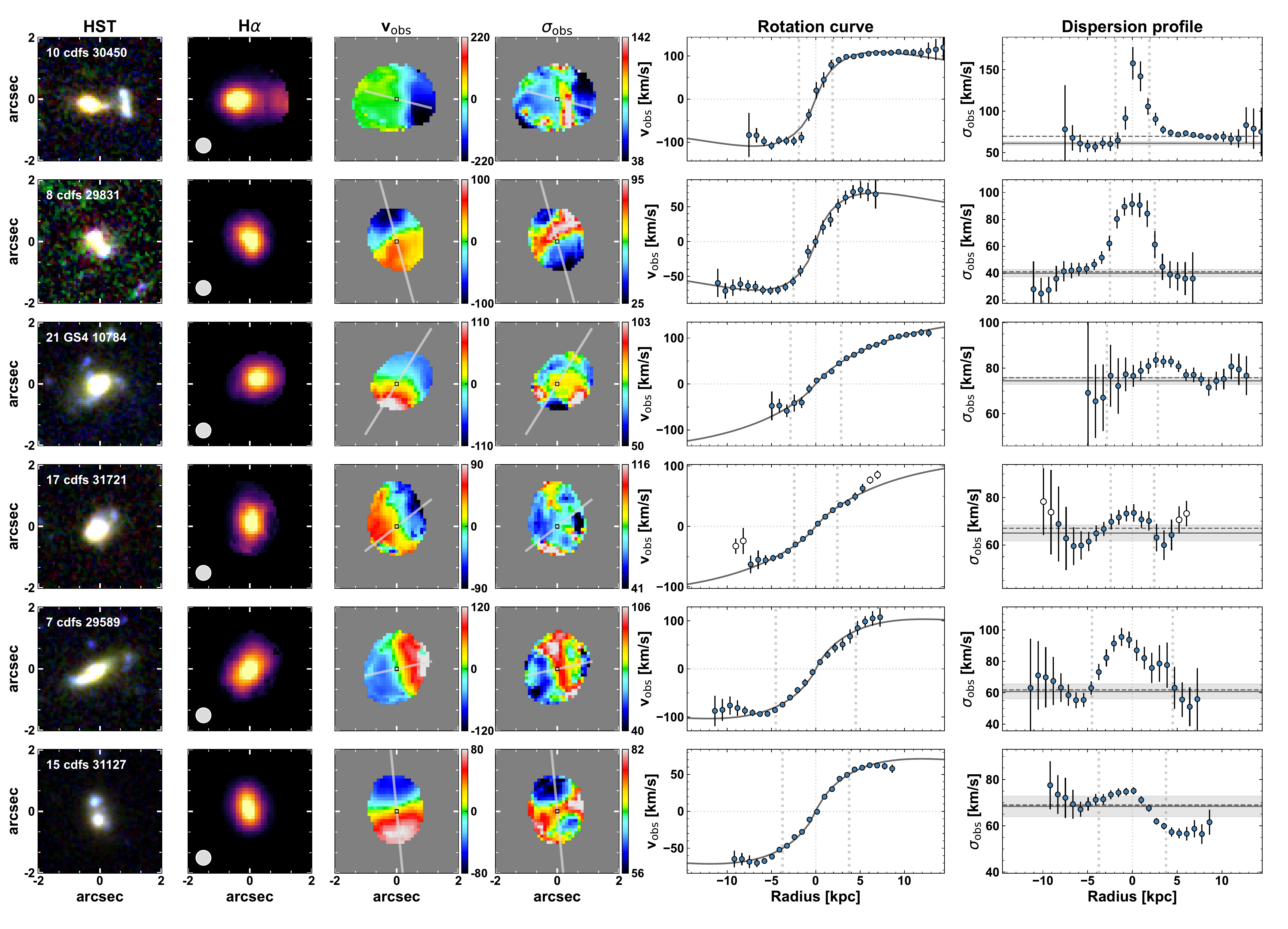}
\caption{Same as Figure \ref{fig:kurvs}.}
\label{figA1:kurvs_dataset2}
\end{figure*}

\begin{figure*}
\includegraphics[angle=90,origin=c,scale=0.45]{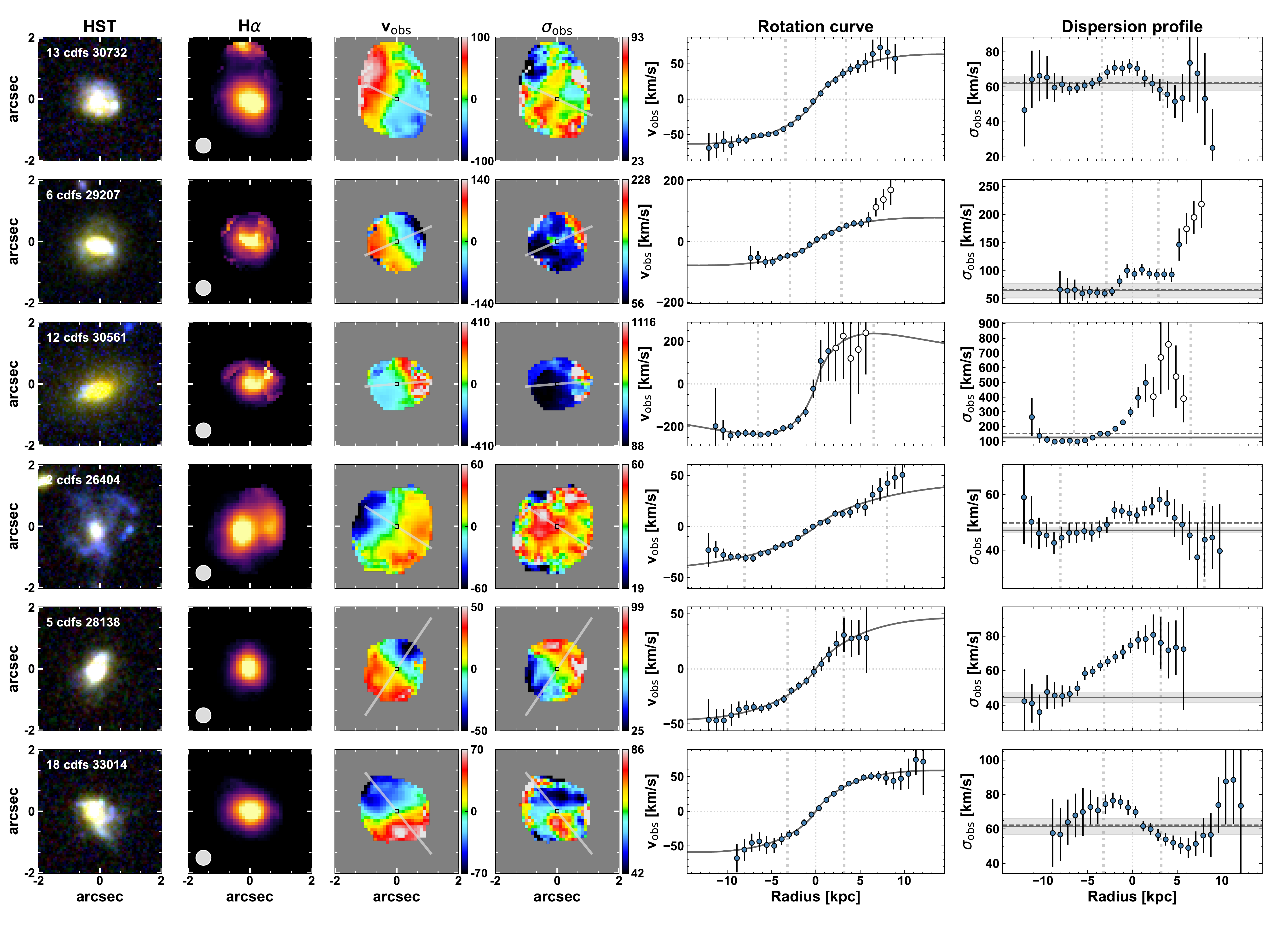}
\caption{{\it continuation}.}
\label{figA2:kurvs_dataset3}
\end{figure*}

\begin{figure*}
\includegraphics[angle=90,origin=c,scale=0.45]{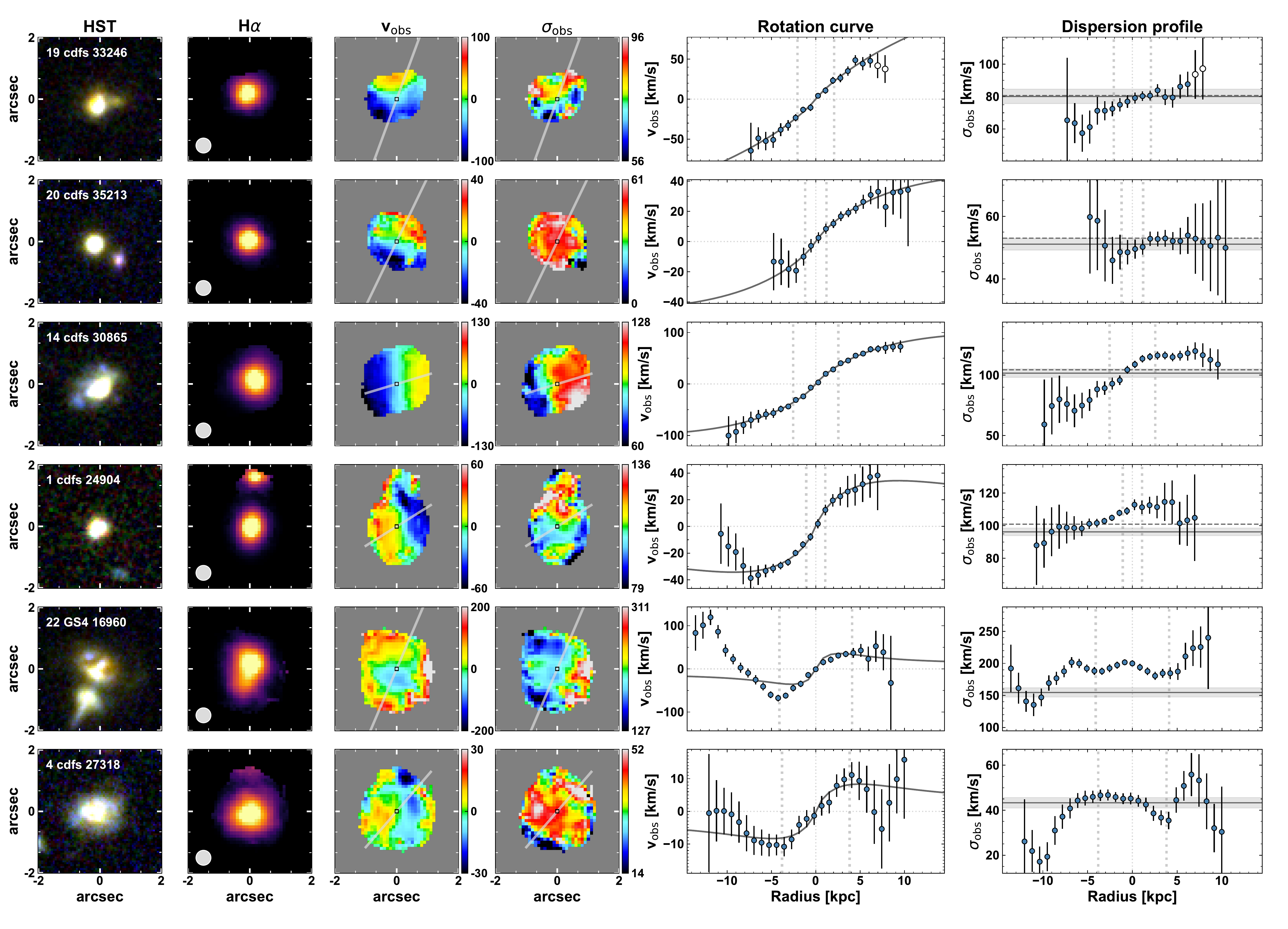}
\caption{{\it continuation}.}
\label{figA3:kurvs_dataset4}
\end{figure*}

\begin{figure*}
\includegraphics[angle=90,origin=c,scale=0.43]{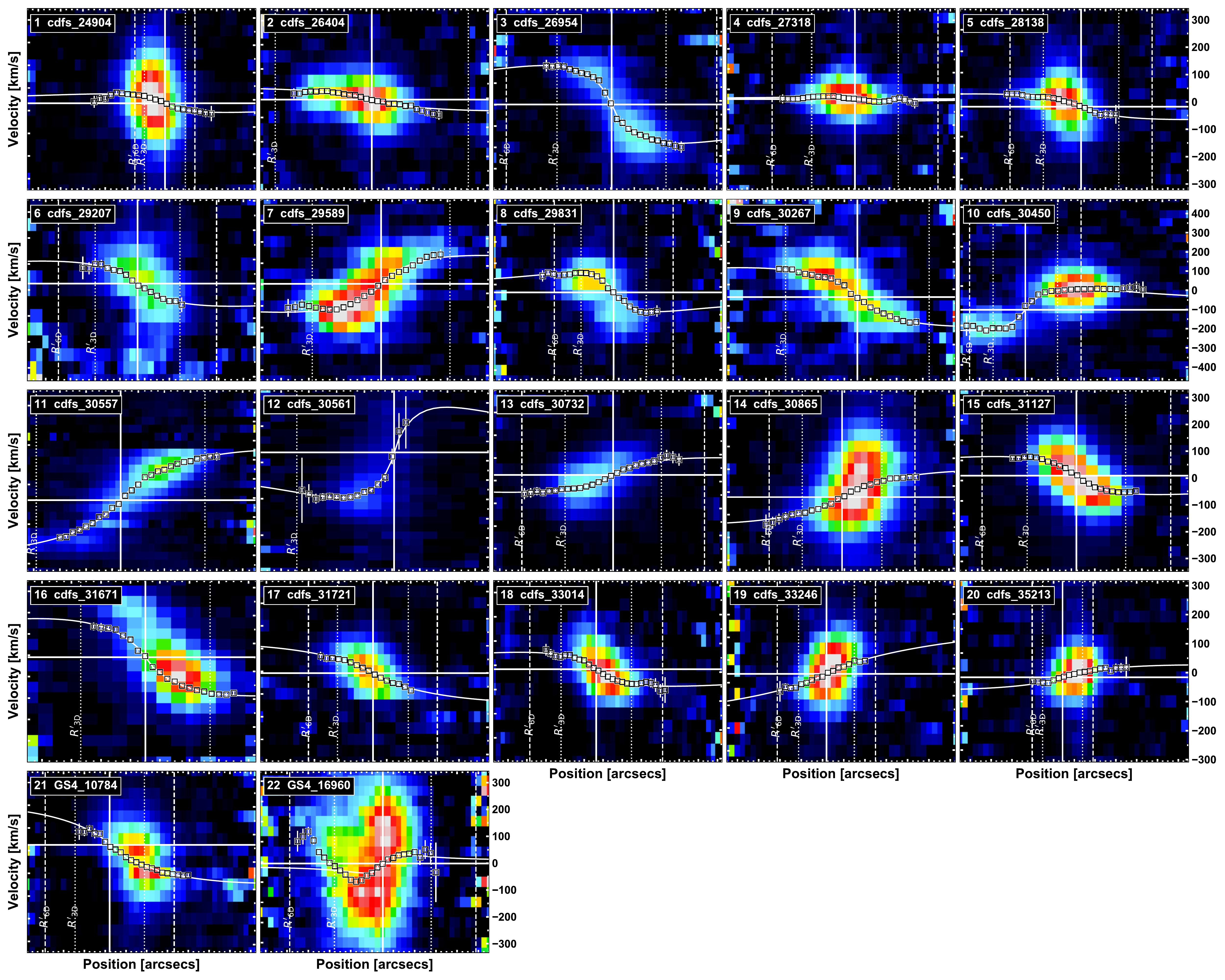}
\caption{ Position-velocity diagrams for the KURVS galaxies. The colour scale represents the flux intensity, with black and white indicating the lowest and highest flux levels, respectively. The colour scale is weighted by the inverse of the sky spectrum to the fifth power.
Open black squares indicate the observed rotation velocity at each pixel extracted from the velocity map and used to obtain the best-fitting exponential disc model (white solid curve). 
Solid vertical and horizontal lines indicate the radial and velocity position of the rotation curve centre, respectively. White, dotted (dashed) vertical lines are displayed at $\pm 3$ times ($\pm 6$ times) the disc scale radius $R_{\rm d}$. 
{A visual inspection of this plot suggests that the best-fitting exponential disc model provide a good extrapolation of the position-velocity diagrams at large radii.}}
\label{fig:kurvs_pv}
\end{figure*}

\subsection*{Notes on individual sources}

\begin{itemize}
\item KURVS-1 or cdfs\_24904. 
This is a compact, nearly spheroidal galaxy, as suggested by the {\it HST} high-resolution imaging and measurements of the near-infrared half-light radius. 
KURVS ultra-deep observations show a somewhat regular velocity field in the inner regions of the galaxy as well as some asymmetries at large radii possibly suggesting the presence of a secondary component. The galaxy shows high velocity dispersion and dispersion-dominated dynamics.

\item KURVS-2 or cdfs\_26404.
The {\it HST} imaging indicates that this is a large clumpy spiral galaxy nearly face-on. Clumps are also observed in the H$\alpha$ flux map. 
KMOS observations indicate a regular velocity field with some asymmetries at both edges of the rotation curve. However, the outer regions of the rotation curve are associated with large uncertainties, and it is difficult to establish if these asymmetries are tracing perturbations at large radii. 
The galaxy is overall regularly rotating, but we exclude it from the ``discs sample'' because of its $v_{\rm rot}/\sigma_0$ ratio.

\item KURVS-3 or cdfs\_26954.
This is a massive spiral galaxy with a prominent bulge in the {\it HST} imaging. 
The galaxy has a regular velocity field and centrally-peaked velocity dispersion profile, suggesting rotationally-supported dynamics. This is also reflected in the $v_{\rm rot}/\sigma_0$ ratio, which is well above unity for this galaxy. 
The galaxy shows a declining rotation curve at $\sim 3 \sigma$ significance. 

\item KURVS-4 or cdfs\_27318.
This spiral galaxy shows a bright bulge in the {\it HST} imaging. The {\it HST} image also shows indication of a possible companion at large radius, beyond the extent of the cut-out shown in Figure \ref{figA3:kurvs_dataset4}. 
The velocity map and velocity dispersion are highly perturbed, and the small velocity gradient indicate that the galaxy is nearly face on. 
The highly-perturbed nature of this galaxy is also indicated by $v_{\rm rot}/\sigma_0 \sim 0.2$. 
The rotation curve of this galaxy is significantly declining at large radii. However, measurements of $t$ are highly uncertain in this galaxy as a result of its very perturbed kinematics.

\item KURVS-5 or cdfs\_28138.
This is a clumpy galaxy with a regular but asymmetric velocity field. The galaxy also shows an asymmetric velocity dispersion profile. 
We exclude it from the ``discs sample'' because of its $v_{\rm rot}/\sigma_0$ ratio.

\item KURVS-6 or cdfs\_29207. 
The {\it HST} morphology suggests a low-inclination spiral galaxy with a bright central component and spiral arms. 
The velocity map indicates regular rotation, and the velocity dispersion map is centrally peaked. 
There are a few ``noisy'' pixels at the edge of both the rotation curve and the velocity dispersion profile. 
From a visual inspection of the KMOS cube, these pixels seems to be associated with a broad component having high velocity dispersion.
Excluding these pixels from the fit to the rotation curve does not change the outer rotation curve shape. 
The galaxy is not included in the  ``discs sample'' because of its $v_{\rm rot}/\sigma_0$ ratio.

\item KURVS-7 or cdfs\_29589. 
This is a highly-inclined spiral galaxy. 
The velocity field suggests regular rotation, although the rotation curve is somewhat asymmetric.
The velocity dispersion profile is centrally peaked. 
This suggests that the galaxy is rotationally-supported in its dynamics, as also indicated by the $v_{\rm rot}/\sigma_0$ ratio.

\item KURVS-8 or cdfs\_29831.
The {\it HST} images indicate a small, clumpy galaxy.
KURVS observations and measurements of  the $v_{\rm rot}/\sigma_0$ ratio indicate rotationally-supported dynamics, as suggested by a regular velocity field and centrally-peaked velocity dispersion profile. 
The galaxy shows a declining outer rotation curve at $\sim 3 \sigma$ significance. 

\item KURVS-9 or cdfs\_30267.
This is a clumpy spiral galaxy with a bright central component in the {\it HST} imaging. 
The regular velocity field and centrally-peaked velocity dispersion suggest rotation-dominated dynamics. 

\item KURVS-10 or cdfs\_30450. {This galaxy is formally classified as a rotationally-supported disc as it displays a regularly-rotating velocity field, a continuous velocity gradient along the kinematic major axis, has $v_{\rm rot}/\sigma_{0} \geqslant 1.5$ and the position of the steepest velocity gradient coincides with the peak of the velocity dispersion.
However, the {\it HST} imaging indicates the presence of two merging companions. 
This suggests that the kinematics signatures observed in KURVS data are associated with the orbital motion of a merging pair, rather than with a rotating disc. 
This is expected, given that that there is a high probability of misclassifying merging pairs as discs in seeing-limited observations \citep{Simons19, Rizzo22}.
The highly asymmetric rotation curve as well as the high velocity dispersion measured in the inner regions of the velocity dispersion map support the merging pair interpretation.} 
Because of the complexity associated with modelling a merging pair, we exclude this galaxy from analysis of the dark matter fractions.

\item KURVS-11 or cdfs\_30557.
This is a large clumpy, edge-on spiral with regular kinematics and centrally-peaked velocity dispersion profile. 

\item KURVS-12 or cdfs\_30561.
This is a red, compact galaxy in the {\it HST} images. 
The rotation curve and velocity dispersion profile are highly asymmetric. 
A visual inspection of the KMOS cube suggests that the outer regions of the kinematic maps are associated with a broad component. 
Excluding these pixels from the fit to the rotation curve does not change the outer rotation curve shape. 
The galaxy is not included in the  ``discs sample'' because of its clearly perturbed kinematics, which also implies a $v_{\rm rot}/\sigma_0$ ratio below our ``discs sample'' cut.
The rotation curve of this galaxy is significantly declining at large radii. However, measurements of $t$ are highly uncertain in this galaxy as a result of its very perturbed kinematics.

\item KURVS-13 or cdfs\_30732.
This galaxy is a clumpy spiral according to the {\it HST} images. 
The source displays a regular velocity field and somewhat asymmetric rotation curve. 
The velocity dispersion profile is peaked at the centre, and noisy and irregular towards the outer regions. However, the outer velocity dispersion profile is consistent with being flat within the uncertainties.
The velocity and velocity dispersion maps suggest a rotating disc, and the galaxy is included in the ``discs sample'' given its $v_{\rm rot}/\sigma_0$ ratio.

\item KURVS-14 or cdfs\_30865. 
This is a clumpy galaxy with a regular velocity field and rotation curve.
The galaxy also shows an asymmetric velocity dispersion profile, and a high velocity dispersion.
As a result, the galaxy is not included in the ``discs sample''.

\item KURVS-15 or cdfs\_31127.
A clumpy galaxy with a regular velocity field and symmetric rotation curve.
The velocity dispersion profile is peaked at the centre, although it shows some asymmetries in the outer regions.
The kinematic maps suggest a rotation-dominated disc, consistent with the $v_{\rm rot}/\sigma_0$ ratio.

\item KURVS-16 or cdfs\_31671.
The {\it HST} imaging indicates a clumpy, edge-on disc.
The galaxy shows a regular velocity field and rotation curve, and a centrally peaked velocity dispersion profile.
This suggests that the galaxy is rotationally-supported in its dynamics, as also indicated by the high $v_{\rm rot}/\sigma_0$ ratio.
From a visual inspection of the KMOS cube, there are a few ``noisy'' pixels at the edge of the rotation curve and the velocity dispersion profile. 
Excluding these pixels from the fit to the rotation curve does not change the outer rotation curve shape. 

\item KURVS-17 or cdfs\_31721.
The {\it HST} imaging indicates a clumpy galaxy.
The object shows a regular velocity field and rotation curve, with a few ``noisy'' pixels at the edge. 
Clipping these pixels does not significantly affect measurements of the the outer rotation curve shape. 
The velocity dispersion profile shows a central peak and some asymmetries towards the outer regions. However, the outer pixels have large uncertainties.

\item KURVS-18 or cdfs\_33014.
The {\it HST} morphology suggests an irregular galaxy.
The velocity map indicates overall rotation. However, the outer rotation curve displays symmetric disturbances.
The object presents a perturbed velocity dispersion map and an asymmetric velocity dispersion profile. 
The perturbations in the outer kinematics might indicate  interactions with a companion. 
The galaxy is excluded from the ``disc sample'' because of the relatively low $v_{\rm rot}/\sigma_0$ ratio.

\item KURVS-19 or cdfs\_33246.
The {\it HST} image shows a small, clumpy galaxy.
KMOS observations indicate an asymmetric velocity field and velocity dispersion profile. 
The galaxy has a high velocity dispersion. 
Excluding ``noisy'' pixels at the edge of the rotation curve and the velocity dispersion profile does not significantly change the outer rotation curve shape. 
The galaxy is excluded from the ``disc sample'' because of its low $v_{\rm rot}/\sigma_0$ ratio.

\item KURVS-20 or cdfs\_35213.
This is a compact galaxy, as suggested by the small half-light radius. 
The {\it HST} image indicates two round-shaped components with low inclinations, it is unclear if these are interacting or are tracing clumpy emission.
The object shows a highly asymmetric rotation curve and an irregular velocity dispersion profile, nearly flat. 
The kinematic properties of the source might also indicate merging activity.
The irregular nature of this galaxy is also reflected in the low $v_{\rm rot}/\sigma_0$ ratio.

\item KURVS-21 or GS4\_10784.
The {\it HST} image indicates a bulgy galaxy with blue emission at the edges. It is unclear if this emission traces spiral arms or disturbances.
The rotation curve and velocity dispersion profile are asymmetric.
There are issues with measurements of the radial and velocity offset obtained from fitting the 1D rotation curve.
The object is formally rotation-dominated given the high $v_{\rm rot}/\sigma_0$ ratio, and we include it in our `discs sample'. However, its kinematic properties suggest perturbations.
Indeed, it is difficult to model the velocity field with \textsc{galpak}$^{\rm 3D}$, and this is independent from measurements of the dynamical center. 

\item KURVS-22 or GS4\_16960.
The {\it HST} morphology is highly perturbed, as well as the velocity and velocity dispersion maps.
As discussed in the main text, the galaxy is detected in X-rays and in the far-infrared, possibly indicating that an AGN and/or a merger-driven starburst are powering its H$\alpha$ emission.

\end{itemize}

\section{Circular velocity and outer rotation curve slopes of $z \sim 0$ discs}
\label{A2:RCs_z0}

We extract one-dimensional rotation curves for THINGS galaxies by applying a similar procedure to that used on the KURVS sample. 
We obtain individual velocity maps from the THINGS webpage\footnote{\url{https://www2.mpia-hd.mpg.de/THINGS/Data.html}}. We extract rotation curves along the kinematic position angle indicated in table 1 of \citet{Trachternach08} and within a 2 pixel radius pseudo-slit. 
We measure the relevant kinematic parameters (i.e., the slope of the rotation curve $t$ and the velocity at $R_{\rm 6D}$) from the observed rotation curve smoothed with a uniform filter with a 100~km~s$^{-1}$ window. This is to mitigate the effect of local disturbances in the rotation curve. 
We checked, however, that measurements of the $t$ parameter for this sample are not affected by the size of the smoothing window. 
We exclude from the plot the four THINGS galaxies with an observed rotation curve that does not extended to $R_{\rm 6D}$.
For the \citet{Catinella06} sample we measured rotation curve slopes from the best-fitting one-dimensional models to their rotation curve templates in bins of I-band magnitude.

We do not apply any pressure-support corrections to measure circular velocities in galaxies in the local Universe. 
This is because the velocity dispersion is much lower in $z \sim 0$ galaxies than in the high-redshift Universe \citep[e.g.][]{Ubler19}, hence the pressure support would correspond to a small correction to the rotation velocity.

\bsp	
\label{lastpage}
\end{document}